\documentclass[a4paper,fleqn, nologo]{cas-sc}

\ExplSyntaxOn
\RenewDocumentCommand \printemails { }
{
  \group_begin:
  \int_compare:nNnTF { \int_use:N \g_ead_int } > { 0 }
  {
    \tex_let:D \thefootnote \relax \footnotetext
    {
      \raggedright
      \textit{Email~address:\c_space_token}
      \seq_use:Nn \g_stm_ead_seq { ;~ }
    }
  }
  { }
  \group_end:
}
\ExplSyntaxOff

\usepackage[numbers]{natbib}
\usepackage{gensymb}
\def\tsc#1{\csdef{#1}{\textsc{\lowercase{#1}}\xspace}}
\tsc{WGM}
\tsc{QE}
\usepackage{xcolor}
\usepackage{soul}
\usepackage{float}

\usepackage{hyperref}

\usepackage{fancyhdr}
\usepackage{lastpage}
\newcounter{video}
\renewcommand{\thevideo}{S\arabic{video}}
\newcommand{\videolabel}[1]{\refstepcounter{video}\label{#1}Video~\thevideo}

\usepackage[pagewise]{lineno}

\definecolor{lightyellow}{HTML}{FFF4CF}
\definecolor{lightblue}{HTML}{D2E2F2}
\definecolor{lightpink}{HTML}{F0CDCC}

\begin{document}
\let\WriteBookmarks\relax
\def\floatpagepagefraction{1}
\def\textpagefraction{.001}

\shorttitle{}    

\shortauthors{}  

\title [mode = title]{One-Shot Generative Design for Disordered Metamaterials via Self-Organizing Neural Cellular Automata}  



%

\author[1]{Yujie Xiang}



\credit{Investigation, Methodology, Validation, Visualization, Writing – original draft, Writing – review \& editing}
\affiliation[1]{organization={Department of Mechanical Engineering, Carnegie Mellon University},
            addressline={5000 Forbes Avenue}, 
            city={Pittsburgh},
            postcode={15213}, 
            state={PA},
            country={USA}}
            
\author[1]{Liwei Wang}


\cormark[1]

\ead{liweiw@andrew.cmu.edu}


\credit{Conceptualization, Funding acquisition, Supervision, Writing – original draft, Writing – review $\&$ editing}

\cortext[1]{Corresponding author}



\begin{abstract}
Disordered metamaterials feature microstructures with inherent randomness and irregularity, enabling them to achieve broader property coverage and superior performance unavailable in their regular counterparts. Despite their promise, designing disordered microstructures is substantially harder than designing regular ones. Their design remains trapped between manual parameterizations with limited expressiveness, and generative AI that is data-hungry and struggles to generalize. To address these limitations, we propose a generative design framework based on Neural Cellular Automata that dynamically grows complex microstructures through learned local interaction rules, inspired by the self-organizing processes in natural materials. This framework requires only a single training template, yet accommodates diverse disordered microstructures and adapts to irregular domains and arbitrary discretizations. By manipulating the learned local rules, we can steer the growth process to generate microstructures unseen during training, providing control over orientation, anisotropy, and directional thickness without retraining. As a dynamic, local growth process, it naturally produces spatially varying microstructures that transition smoothly to enable location-specific mechanical properties. We demonstrate this in a multiscale mechanical cloaking design, where microstructures vary across the space to meet an optimized heterogeneous property distribution. Our design enables excellent cloaking performance without complicated post-processing and incompatible assembly common in existing methods. This data-efficient, generalizable approach opens access to previously intractable disordered materials for biomedical implants and soft robotics.

\end{abstract}









\maketitle 
\thispagestyle{fancy}
\section{Introduction}\label{}
Mechanical metamaterials derive their properties mainly from the geometry of their microstructures~\cite{Jiao2023, Bertoldi2017, lee2024data}. They can achieve mechanical behaviors inaccessible to conventional materials~\cite{Buckmann2014, Boddapati2024, Lin2024}, such as negative Poisson's ratio~\cite{Zhang2023}, highly tunable energy absorption~\cite{Shaikeea2022, Surjadi2026}, and superior impact resistance~\cite{Liu2022}. With their excellent performance and tunability, mechanical metamaterials have transformed applications ranging from lightweight structures to soft robotics~\cite{Gregg2024}, acoustic control~\cite{Cummer2016,wang2022generalized}, and biomedical devices~\cite{MetaScaffold2025}. Most existing mechanical metamaterials assume ordered microstructures, featuring deterministic and regular arrangements of predefined building blocks~\cite{Jiao2023, Bertoldi2017, WangCloak2022, Feng2026}. This order in assembly makes the building-block microstructures and their multiscale systems easy to parameterize and design~\cite{Pahlavani2024, Fang2022}. However, it also constrains the design space, limiting access to heterogeneous and more morphologically diverse microstructures~\cite{Zaiser2023}. Natural materials, such as trabecular bone~\cite{Oftadeh2015}, sea sponge~\cite{WangSponge2023}, coral~\cite{Shen2026}, and nacre~\cite{Gim2019}, are instead built on disordered microstructures~\cite{Luan2023}, featuring randomness, irregularity, and interconnectivity.  Such structural disorder helps suppress deformation localization~\cite{LiuRobustness2024}, deflect crack propagation~\cite{Shaikeea2022, Fulco2025a, Fulco2025b}, redistribute stress~\cite{Jia2024}, and dissipate impact energy~\cite{Liu2025}, leading to greater robustness and performance unavailable in ordered structures. These unique advantages reveal only the tip of a much broader design space and performance envelope hidden within disorder~\cite{Zaiser2023, Reid2018, Hanifpour2018}.

Despite their superior performance, disordered microstructures are more difficult to design than ordered ones, due to their inherent randomness and irregular geometries. A growing number of computational methods have emerged to address this challenge~\cite{Zaiser2023, Dhulipala2025, Xu2014a, Xu2014b, Deng2026}. One class of approaches generates disordered microstructures through prescribed rules or basis functions, including Voronoi tessellations that partition the domain around randomly placed seed points~\cite{Zheng2023}, Gaussian random fields or trigonometric basis functions to create spinodal-like bicontinuous microstructures~\cite{He2024, Kumar2020, Zheng2021, Senhora2022, Deng2025, Portela2020}, and wave function collapse methods that tile local pattern fragments to assemble globally consistent textures~\cite{Liu2022, Jia2024, Jia2024b}. These rule-based methods address the parameterization challenge of disordered structures, yet each is confined to a narrow morphological class constrained by its construction rule, with no straightforward way to go beyond, transition between, or mix different microstructure types. A substantial portion of the vast design space of disordered materials thus remains untapped. Moreover, they require human-specified rules, which are nontrivial to obtain for the diverse, highly complex microstructures found in nature. However, these microstructures are precisely the ones that have great potential for exceptional properties and niche applications. 

Deep generative models offer an alternative. Variational autoencoders~\cite{WangGenerative2020, Xiang2025}, generative adversarial networks~\cite{ShenBuehler2022, Mao2020}, and diffusion models~\cite{Bastek2023, Guo2024} learn the statistical distribution directly from disordered samples to reproduce microstructures without explicitly specified rules, which is more flexible than rule-based methods. However, this capability introduces several new limitations. First, training these models typically requires a large amount of microstructure samples, which are difficult to acquire~\cite{QuesadaMolina2025}. Second, their ability to generalize beyond the training distribution is limited, as they struggle to generate morphologies that are underrepresented in the training data, or truly novel ones unseen in their training data~\cite{ShenBuehler2022}. Third, the trained models are also strictly tied to the domains and discretizations on which they were trained (oftentimes a square or cube with a uniform grid) and transfer poorly to other settings, typically requiring complete retraining. 

In summary, rule-based methods can transfer across domains and discretizations but are restricted to manually prescribed rules, while deep generative models directly learn from data but require large datasets and remain tied to a fixed domain and discretization. Therefore, unlocking the full potential of disordered metamaterials requires a generative framework that can learn from data while remains data-efficient and domain-adaptable~\cite{Jiao2023, Zaiser2023}.

\begin{figure}
  \centering
  \includegraphics[width=1.0\textwidth]{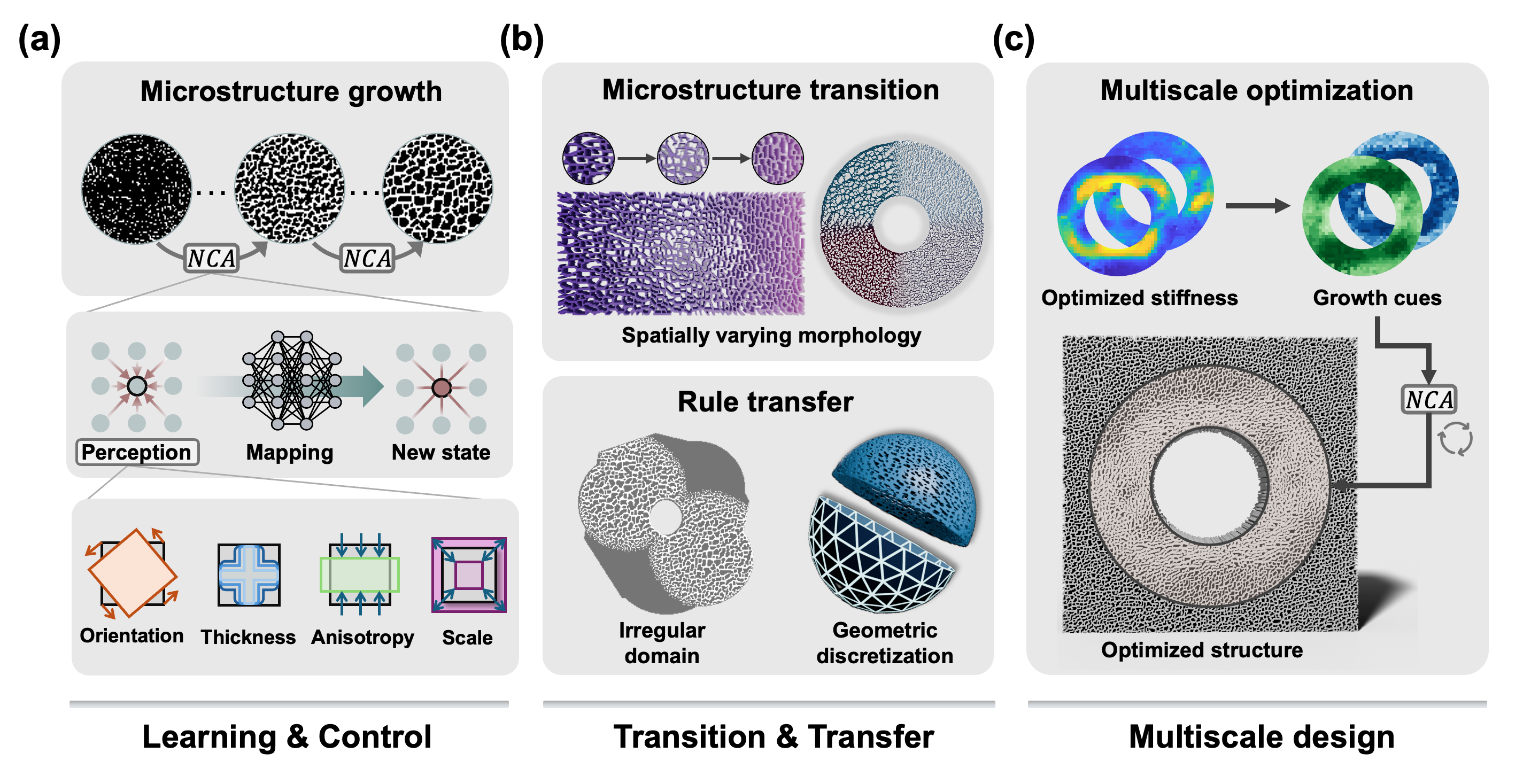}
  \caption{Overview of the proposed generative framework for disordered microstructures. \textbf{(a)}  Learning and controls on homogeneous microstructure. An NCA model is trained to iteratively grow a microstructure through local perception and mapping stages, where each cell perceives its neighborhood and maps to a new state. Four steering mechanisms provide continuous control over the orientation, thickness, anisotropy, and scale of the growth process without any retraining. \textbf{(b)} Transition in heterogeneous microstructure and transfer to different geometries. Heterogeneous microstructure with smooth morphological transitions grow for spatially varying controls and multiple types. The same trained model transfers to different generation settings, including irregular domains and different geometric discretizations, by adapting only the local perception. \textbf{(c)} Multiscale design from microscale to macroscale. The generative framework is coupled with optimization workflow, where the optimized macroscale stiffness distribution is converted into spatially varying growth cues that drive the trained NCA model to grow the full optimized structure.}\label{fig_1}
\end{figure}

Nature, the very thing that inspired the exploration of disordered metamaterials, provides a clue to address this gap. Many disordered microstructures we see in nature are formed by self-organization, where individual cells or chemicals respond only to their neighbors, yet global patterns emerge from these local interactions without any central blueprint.~\cite{Turing1952, Kondo2010}. This self-organizing principle creates microstructures whose diversity and complexity far exceed those of our engineered systems, adapting freely across domains and scales in contrast to our brittle models that cannot generalize. For example, leaf veins form through local transport and reinforcement between neighboring cells, yet these local interactions collectively produce a connected vascular network across the entire leaf with varying shapes~\cite{Runions2005}. This principle inspires us to propose a fundamentally different design paradigm built on Neural Cellular Automata ~\cite{Mordvintsev2020, Seibert2024, Pajouheshgar2026}, which embeds such self-organizing dynamics to generate disordered microstructures (Fig.~\ref{fig_1}). Instead of learning the static microstructures themselves, as existing AI-driven methods do, we aim to find an underlying local interaction rule that dynamically grows them. And instead of constructing such a rule by hand, as in existing rule-based methods, we use neural networks to learn it from samples. In this way, the framework combines the locality of rule-based methods with the expressive power of neural networks. The former makes it data-efficient and generalizable to unseen domains, while the latter allows it to accommodate complex microstructures and different types. With it, we can grow disordered microstructures in a controllable way across scales, domains, and discretizations, just as nature grows its microstructures, but now in silico.

Specifically, we apply this self-organizing principle at three progressively increasing levels of complexity. We first focus on generating homogeneous microstructures (Fig.~\ref{fig_1}(a)). An NCA model, trained from a single template, learns a dynamic growth process in which each cell iteratively extracts neighboring information through local perception and maps it to a state update through a shared network, growing microstructures that match those of the template. We further introduce post-training control mechanisms into the learned dynamics to manipulate orientation, anisotropy, characteristic scale, and directional thickness of generated microstructure. By combining these controls, we expand the mechanical property space far beyond that of the training template~\cite{Andreassen2014}.

We then extend the framework to heterogeneous microstructures across different domains and discretizations (Fig.~\ref{fig_1}(b)). It produces spatially varying microstructures with smooth morphological transitions between different growth controls and microstructure types. By adapting the local perception in the self-organizing rule, we transfer the learned rule to irregular domains and arbitrary discretizations different from the training setting.

Finally, we couple the growth rules with topology optimization to enable multiscale design with optimized system performance(Fig.~\ref{fig_1}(b)). The optimizer assigns spatially varying microstructure properties, which are transformed into spatially varying growth control parameters, akin to growth cues in natural materials. This allows us to dynamically grow the entire multiscale structure using the same learned NCA while following spatially varying growth cues. We use mechanical cloaking design as an extreme case to stress-test the method: it produces smoothly varying, disordered multiscale structures that accommodate local property requirements and deliver excellent cloaking behavior.

Together, this study makes the following contributions:
\begin{itemize}

\item We establish a lightweight one-shot generative framework that learns self-organizing rules from a single sample and grows disordered microstructures across different domains and scales.

\item By treating the NCA as a generalized PDE system, we achieve flexible control of both morphologies and mechanical properties beyond the training template without new data collection or retraining.

\item The learned self-organizing rules transfer across irregular domains and discretizations different from the training setting. Their asynchronous growth further enables scalable generation in large domains, as well as self-repair.

\item We integrate the NCA into a multiscale optimization pipeline, generating heterogeneous multiscale structures that meet location-specific property requirements with smooth transitions between them.

\end{itemize}

The remainder of this paper is organized as follows. 
Section~\ref{sec:nca} introduces the NCA model, its one-shot training strategy, and post-training steering mechanisms for the generation of homogeneous disordered microstructures. Section~\ref{sec:extension} extends the framework to generate heterogeneous microstructures and adapt to a 3D surface mesh. Section~\ref{sec:multiscale} presents the multiscale optimization pipeline and demonstrates its application to the mechanical cloaking problem. Section~\ref{sec:conclusion} concludes the paper with a discussion of limitations and future directions.

\section{One-shot learning for control growth of homogeneous microstructures}\label{sec:nca}

In this section, we establish the NCA-based generative model for disordered microstructure design. We first establish the NCA model and demonstrate its flexibility across distinct disordered microstructure types. We then implement four steering mechanisms to control both morphologies and mechanical properties over the generated microstructure without retraining the model. Finally, we explore how these controls broaden the mechanical property space of the generated microstructures, laying the foundation for the multiscale design in Sec.~\ref{sec:multiscale}.


\subsection{One-shot learning with Neural Cellular Automata}
\subsubsection{NCA models}\label{subsubsection:NCA models}
Neural Cellular Automata (NCA) provide a natural framework for learning self-organizing dynamics to grow microstructures from local interactions. We consider each discretized component in a microstructure as a cell, which can be a pixel, voxel, element, or node. In an NCA, each cell iteratively updates its state based on information perceived from the states of its neighborhood~\cite{Mordvintsev2020}. Instead of manually prescribing the update rule, the model represents the update rule with a neural network shared across all cells, whose parameters are learned directly from data. By repeatedly applying the learned update rule across space and time, complex global morphologies emerge from purely local interactions, making NCA well suited for modeling disordered microstructures with diverse and irregular morphologies.

\begin{figure}
  \centering
  \includegraphics[width=0.85\textwidth]{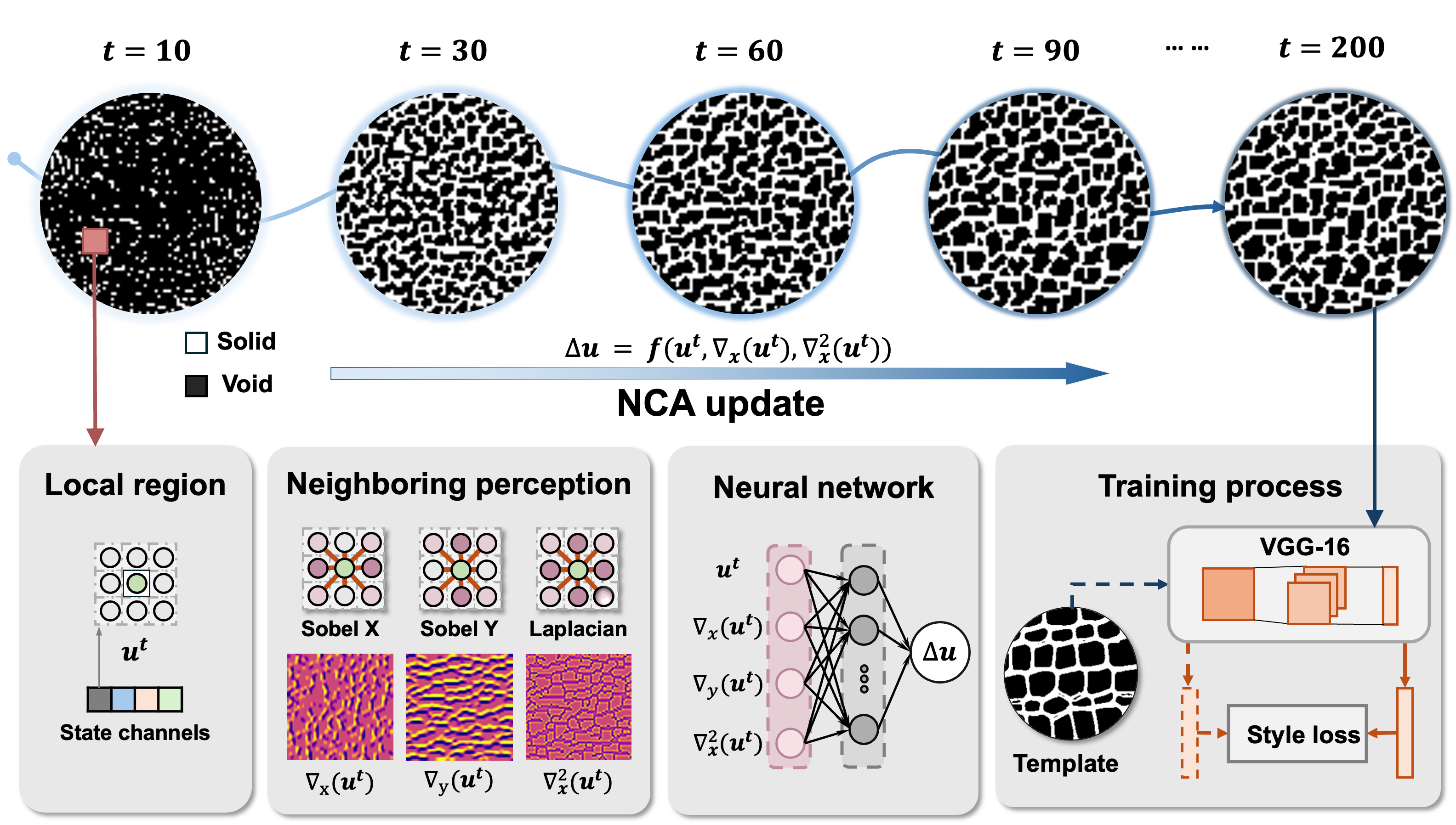}
  \caption{Growth process and model architecture of the NCA framework illustrated on the leaf vein microstructure. Each cell carries a multi-channel state vector $\mathbf{u}^t$, whose first channel encodes the solid-void phase, corresponding to the local regions in the microstructure. The model iteratively applies a shared local update rule through fixed perception and mapping stages to grow the microstructure. Local neighborhood information is extracted by fixed Sobel and Laplacian filters, with the colored cells indicating different weights applied to each neighbor, producing the gradient and Laplacian maps shown below. The neural network for mapping is trained in a one-shot manner from a single template using a frozen VGG-16 perceptual feature extractor, where a sliced optimal transport loss matches the multiscale statistics of the generated microstructure to the template.}\label{fig_2}
\end{figure}

Specifically, each cell in the growth domain carries a state vector $\mathbf{u} \in \mathbb{R}^{C}$, where the first channel stores the solid-void phase and the remaining $C-1$ channels are hidden channels that allow cells to encode and exchange latent information, providing the memory and context needed for coherent global patterns to emerge, as illustrated in the local region and state channels in Fig.~\ref{fig_2}. In each update step, the state of every cell $\mathbf{u}^{(t)}$ is updated according to
\begin{equation}
    \mathbf{u}^{(t+1)}
    = \mathbf{u}^{(t)}
    + \delta^{(t)} \cdot F_{\theta}\!\left(\mathit{percept}(\mathbf{u}^{(t)})\right),
    \label{eq:stochastic_update}
\end{equation}
where $\mathit{percept}$ is a fixed local perception operator to collect neighboring information, $F_{\theta}$ is a trainable neural network shared across cells that maps this neighboring information to an incremental state update, and $\delta^{(t)}$ is a stochastic update mask. The stochastic update mask $\delta^{(t)}$ in Eq.~\eqref{eq:stochastic_update} is sampled independently for each cell from a Bernoulli distribution with $p = 0.5$ at every update step. This stochastic masking breaks spatial symmetry to excite emergent dynamics. It also functions as a form of spatial dropout during training, improving the robustness and generalization of the learned rule to large, irregular, and partially initialized domains, which will be exploited in the asynchronous growth capability introduced in Section~\ref{subsection:asynchronous}.

While the model is general, we first focus on microstructures represented on pixelated regular grids for ease of illustration and extend it to more general representations in later sections. Given a pixelated grid, the local perception operator extracts neighborhood information in each cell through a set of fixed $3\times3$ depthwise convolution kernels: Sobel filters $\mathbf{K}_{x}$ and $\mathbf{K}_{y}$ approximating the spatial gradients $\nabla_x\mathbf{u}$ and $\nabla_y\mathbf{u}$, and a Laplacian filter $\mathbf{K}_{\mathrm{lap}}$ approximating $\nabla^2\mathbf{u}$, as shown in Fig.~\ref{fig_2}. Their outputs are concatenated with the cell's own state to form the perception vector
\begin{equation}
    \mathit{percept} = \bigl[\,\mathbf{u},\;
    \mathbf{K}_{x} \ast \mathbf{u},\;
    \mathbf{K}_{y} \ast \mathbf{u},\;
    \mathbf{K}_{\mathrm{lap}} \ast \mathbf{u}\,\bigr]
    \in \mathbb{R}^{4C}.
    \label{eq:perception}
\end{equation}

\noindent Note that other sets of operators can be chosen for more general settings. We then pass the perception vectors to the shared neural network $F_{\theta}$, which outputs a residual state update $\Delta\mathbf{u} = F_{\theta}(\mathit{percept})$. The same network is shared across all cells, so that each cell executes the same local update rule on different neighbors, keeping the number of trainable parameters small and, crucially, independent of the grid size. A key insight of this study is that the resulting model can be interpreted as a generalized partial differential equation (PDE), allowing us to leverage rigorous operator theory and a rich set of analytical tools to improve controllability and generalization. Specifically, the update rule can be viewed as a forward Euler discretization of the following PDE:
\begin{equation}
    \frac{\partial \mathbf{u}}{\partial t}
    = \mathcal{F}_{\theta}\!\left(
        \mathbf{u},\;
        \nabla \mathbf{u},\;
        \nabla^{2} \mathbf{u}
    \right),
    \label{eq:pde}
\end{equation}
where $\mathcal{F}_{\theta}$ is a nonlinear evolution function acting on the local PDE operators, approximated by $F_{\theta}$ on locally perceived information in Eq.~\eqref{eq:perception}. This PDE perspective unifies the NCA with classical self-organizing systems such as reaction-diffusion models~\cite{Turing1952, Kondo2010}. In both cases, the state $\mathbf{u}$ evolves through purely local interactions represented by the local perception operator. Both also share the same emergent character: coherent global patterns arise from the cumulative effects of repeated local interactions across the domain. The key distinction is that classical PDE systems rely on manually prescribed rules for $\mathcal{F}$, which confines them to a narrow class of morphologies, whereas the NCA learns $\mathcal{F}_{\theta}$ directly from data, replacing hand-crafted rules with a trainable approximation that can capture the complex local dynamics to produce any target morphology. We can also view classical PDE systems as special cases of NCA with a prescribed (e.g., polynomial) form of the evolution operator $\mathcal{F}_{\theta}$. Because the learned rule relies only on local interaction and does not depend on the resolution, representation, or geometry of the training domain, it can be applied to any target resolution or domain with the same network parameters. This reflects a fundamental distinction from conventional deep generative models: rather than learning a static mapping from the latent distribution to an entire structure, the NCA learns the evolution dynamics that grow the microstructure progressively, with the global pattern emerging from cumulative local interactions over time.

\subsubsection{Model training}
Instead of requiring a large dataset of microstructure samples, the NCA update network is trained in a one-shot manner from a single template. The goal of the training is not to replicate each individual pixel in the template, but to learn the local update rule that can grow statistically equivalent morphology~\cite{Bostanabad2016, Bostanabad2018}, which is more meaningful for disordered microstructures as we are learning the universal growth rules rather than a single realization of the template. Therefore, we use the style loss in the training, which measures similarity through feature distributions across multiple spatial scales, allowing positional variation while enforcing statistical similarity. We adopt a frozen ImageNet-pretrained VGG-16 as a fixed feature extractor and apply a sliced optimal transport loss to compute $\mathcal{L}_{\mathrm{style}}$ by matching the feature distributions of the generated microstructure to those of the template across multiple intermediate layers without enforcing any spatial correspondence~\cite{Simonyan2015}. The total training loss combines the style loss $\mathcal{L}_{\mathrm{style}}$ with a state overflow regularization term:
\begin{equation}
    \mathcal{L} = \mathcal{L}_{\mathrm{style}} 
    + \sum_{x,y,c} \bigl| u_{x,y,c} 
    - \mathrm{clamp}(u_{x,y,c},\,-1,\,1) \bigr|,
    \label{eq:total_loss}
\end{equation}
where ${clamp}(\cdot,\,-1,\,1)$ limits each state value to the range $[-1,1]$. The second regularization term therefore penalizes any state value that exceeds $[-1,1]$ to keep the state values bounded and stabilize training over long update sequences. The pretrained VGG-16 parameters are frozen during the training, which is only used for loss calculation. Parameters $\theta$ in the NCA update network $F_{\theta}$ are the only trainable parameters. They are optimized with the Adam optimizer with a stepwise decaying learning rate, with the key hyperparameters summarized in Table~\ref{tbl_a1}.

\begin{figure}
  \centering
  \includegraphics[width=0.85\textwidth]{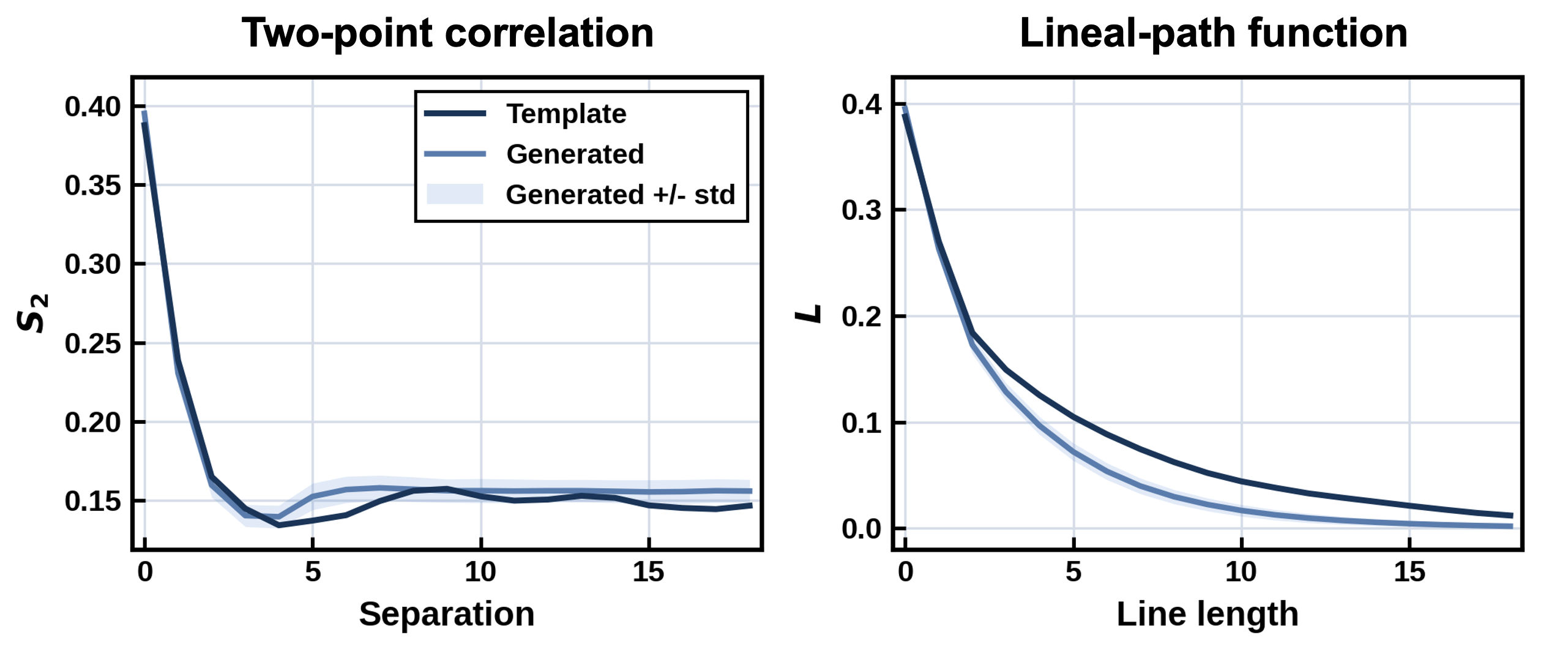}
  \caption{Comparison between the leaf template and generated samples on the two-point correlation function $S_2$, and the lineal-path function $L$. The close overlap between the generated average curves and the template curves, together with the narrow standard-deviation bands, shows that the generated leaf structures reproduce the key morphological features of the template, including spatial correlations and solid phase connectivity.}\label{fig_3}
\end{figure}

With these setups, we validate the framework by training it on a leaf vein microstructure, with the growth process illustrated in Fig.~\ref{fig_2} (see Vid.~\ref{vid:nca growth} for the whole process). Starting from an empty seed state, scattered local fragments first emerge through the learned update rule and gradually connect into a coherent network, continually refining until the morphology stabilizes. The resulting microstructures closely mimic the template. To provide further quantitative validation, we apply two-point correlation and lineal path functions~\cite{Jiao2007} to quantify the morphological features between the template and the generated microstructures in Fig.~\ref{fig_3}. The two-point correlation function measures the probability that two points separated by a given distance both lie in the solid phase, capturing the volume fraction and characteristic length scale, while the lineal path function measures the probability that an entire line segment of a given length lies within the solid phase, characterizing the connectivity of the solid network. The close agreement between the generated microstructures and the template across both metrics confirms that the learned update rule has successfully reproduced the key morphological features and underlying distribution of the template, not merely a single pixel-level realization.

\subsubsection{Validation on multiple types of disordered microstructure}
\begin{figure}
  \centering
  \includegraphics[width=0.85\textwidth]{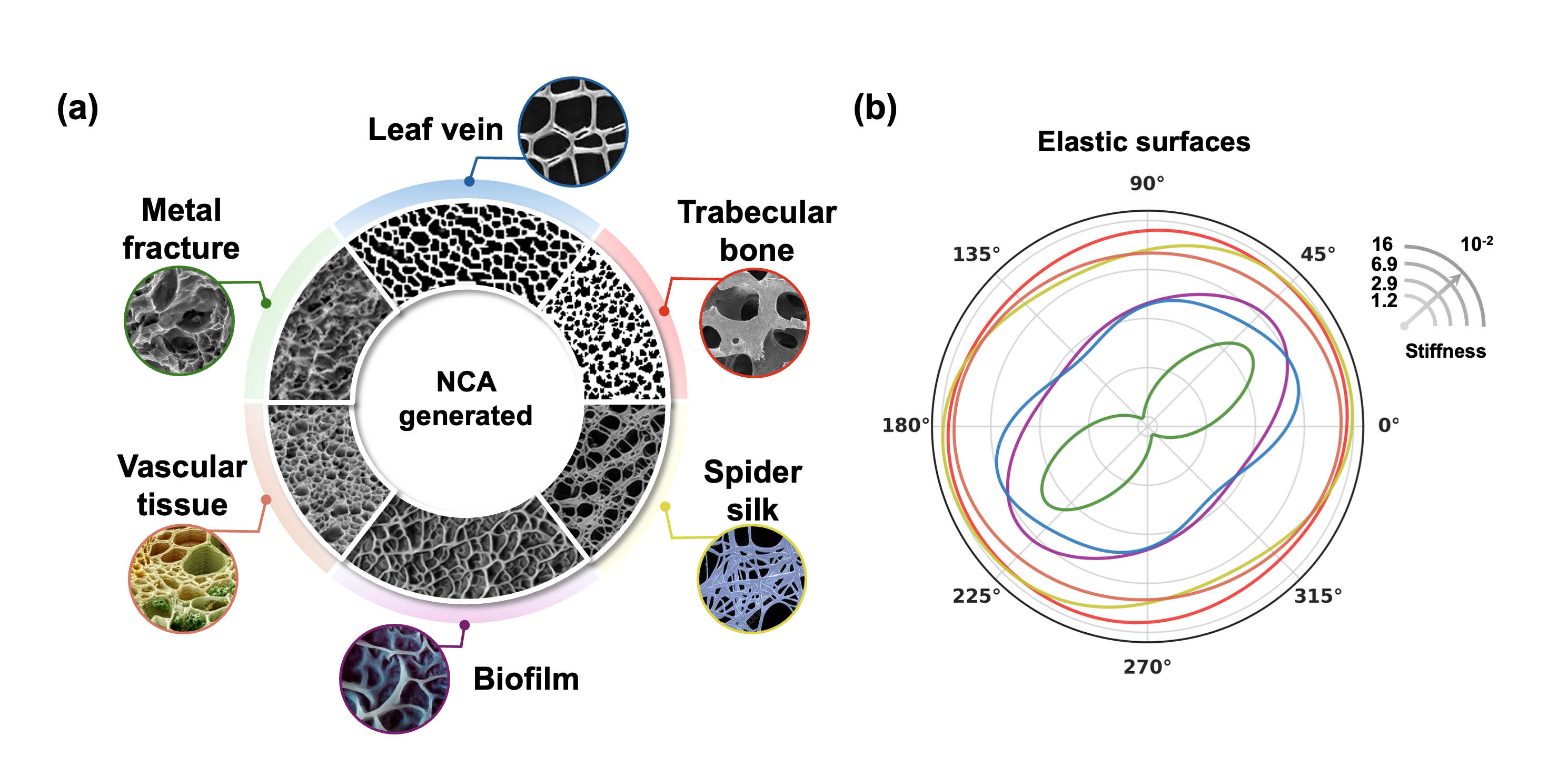}
  \caption{Validation of the NCA generative framework on six biologically and physically distinct disordered microstructures. \textbf{(a)} Six template microstructures (outer ring) and their corresponding NCA-generated realizations (inner ring), each produced by an independently trained model with the same model setups. \textbf{(b)} Homogenized elastic surfaces of the six generated microstructures, plotted in polar coordinates with the radial distance representing the directional stiffness magnitude and the overall shape characterizing mechanical anisotropy.}\label{fig_4}
\end{figure}

To demonstrate the flexibility of our framework, we selected multiple different types of disordered microstructures with distinct morphological features (Fig.~\ref{fig_4}): homogeneous and hierarchical network topologies, as in the leaf vein and vascular tissue; porous architectures with rounded strut junctions, as in the trabecular bone; dendritic branching systems, as in the biofilm; sparse fiber networks characterized by long-range connections and local conjunctions, as in the spider silk; and granular microstructures with sharp interfacial boundaries, as in the metal fracture surface. We trained a separate model for each template using the same architecture and hyperparameters.

The comparison of generated microstructures and their corresponding templates is shown in Fig.~\ref{fig_4}(a). Each of the generated microstructures maintains the key morphological features in those of the corresponding template. The successful generation of these distinct disordered microstructures demonstrates the flexibility and robustness of our proposed generative framework. The homogenized elastic surfaces of these microstructures are shown in Fig.~\ref{fig_4}(b). The six generated microstructures exhibit a wide range of mechanical responses from nearly isotropic to strongly anisotropic, demonstrating that the morphological diversity of the microstructures directly translates into a broad mechanical property space. This is enabled by the flexibility of neural networks over predefined rule-based methods.

\subsection{Post-training microstructure control}\label{subsection:control}
The model structure and training strategy introduced in the previous section enable the NCA to learn a local update rule that reproduces the morphology of a given microstructure. In this section, we aim to extend this generative capability to go beyond reproduction to create new microstructures with controllable variation in the morphology and corresponding mechanical properties without retraining the model. As noted earlier, the learned self-organizing rule in NCA uses local operators in the same way that PDEs use them to describe pattern evolution. In PDE and computational geometry theory, there is a well-established analysis of how changes in these local operators alter the behavior of the PDE evolution. We can therefore transfer this analysis to the learned NCA model, steering the growth process to control the generated morphology (see Vid.~\ref{vid:individual control}), by modifying these local perception operators while keeping the learned update network fixed. Based on this principle, we introduce a set of post-training steering mechanisms to manipulate orientation, anisotropy, characteristic scale, and directional thickness, expanding the morphological diversity and mechanical property space.

\subsubsection{Orientation control through gradient rotation}
\begin{figure}
  \centering
  \includegraphics[width=0.85\textwidth]{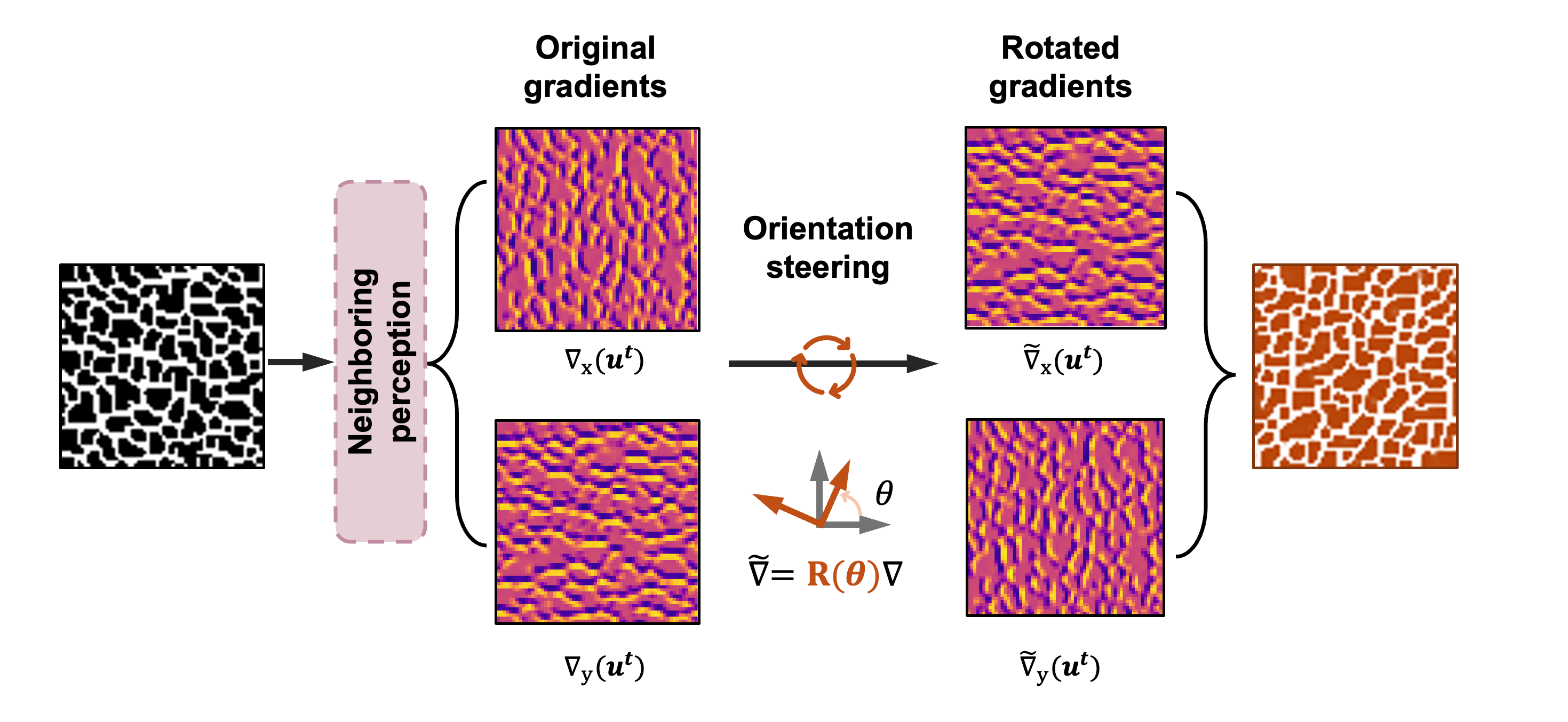}
  \caption{Orientation steering applied to the local perception operation. The native gradients extracted by the Sobel filters are rotated by a prescribed steering angle $\theta$ via a rotation matrix $\mathbf{R}(\theta)$ before being passed to the trained update network, controlling the growth to follow a specific direction.}\label{fig_5}
\end{figure}

\begin{figure}
  \centering
  \includegraphics[width=0.8\textwidth]{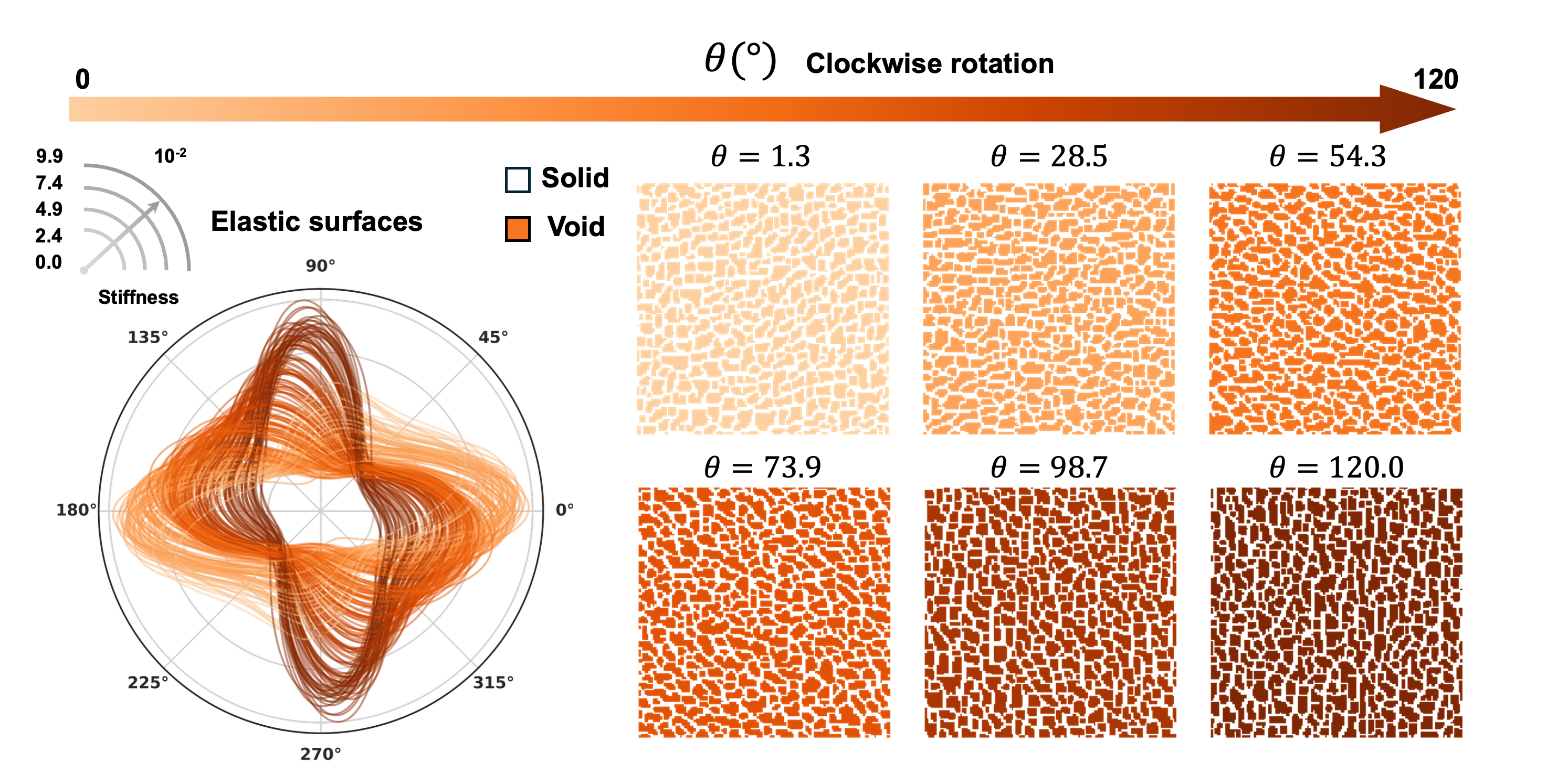}
  \caption{Generated microstructures and corresponding elastic surfaces under orientation steering. An arrow with a gradually changing color shows the increasing orientation angle $\theta$ at the top. The elastic surfaces of 100 independently generated samples at each $\theta$ represent the gradual transition effect in the stiffness space controlled by the steering angle. They confirm that orientation steering rotates the principal stiffness axes without altering the degree of anisotropy. Six examples are illustrated with increasing $\theta$ on the right, demonstrating a morphologically continuous rotation.}\label{fig_6}
\end{figure}

To control the orientation of growth, we apply a rotation matrix $\mathbf{R}(\theta)$ to the spatial gradients in the local perception operation, rotating them by a target angle $\theta$:
\begin{equation}
    \begin{pmatrix} \widetilde{\nabla}_x\,\mathbf{u}^{(t)} \\ \widetilde{\nabla}_y\,\mathbf{u}^{(t)} \end{pmatrix}
    = \mathbf{R}(\theta)
    \begin{pmatrix} \nabla_x\,\mathbf{u}^{(t)} \\ \nabla_y\,\mathbf{u}^{(t)} \end{pmatrix}
    =\begin{pmatrix}
    \cos\theta & \sin\theta \\
    -\sin\theta & \cos\theta
    \end{pmatrix}\begin{pmatrix} \nabla_x\,\mathbf{u}^{(t)} \\ \nabla_y\,\mathbf{u}^{(t)} \end{pmatrix},
    \label{eq:rotated_gradients}
\end{equation}
which can be written compactly as $\widetilde{\nabla} = \mathbf{R}(\theta)\nabla$. As illustrated in Fig.~\ref{fig_5}, these steered gradients replace the original gradients as inputs for the steered update rule
\begin{equation}
    \mathbf{u}^{(t+1)} = \mathbf{u}^{(t)} + \delta^{(t)} \cdot F_{\theta}\!\left(\mathbf{u}^{(t)},\; \widetilde{\nabla}_x\,\mathbf{u}^{(t)},\; \widetilde{\nabla}_y\,\mathbf{u}^{(t)},\; \nabla^2\mathbf{u}^{(t)}\right),
    \label{eq:orientation_update}
\end{equation}

\noindent which provides an orientation-steering mechanism for the self-organizing dynamics, producing microstructures with orientation closely following the prescribed angle (Fig.~\ref{fig_6}). The resulting effective stiffness also rotates accordingly. This gives direct control over both morphology and mechanical anisotropy. Notably, the steering reorients the principal stiffness axes without changing the degree of anisotropy: the elastic surfaces retain nearly the same shape and size across all steering angles.

\subsubsection{Anisotropy and scale control}
Although orientation steering rotates the generated microstructure and its elastic surface, it cannot control the degree of anisotropy. To further control the anisotropy and feature scale, we reshape the local coordinate system in which the NCA extracts neighborhood information using a unified Riemannian metric framework~\cite{Loseille2011}. The steered update rule becomes
\begin{equation}
    \mathbf{u}^{(t+1)} = \mathbf{u}^{(t)} + \delta^{(t)} \cdot F_{\theta}\!\left(
    \mathbf{u}^{(t)},\; 
    \mu^{-1/2}\mathbf{g}^{-1}\nabla\mathbf{u}^{(t)},\; 
    \widetilde{\nabla}^2\mathbf{u}^{(t)}\right),
    \label{eq:anisotropy_scale_update}
\end{equation}
where the Riemannian metric tensor $\mathbf{g}$ and the anisotropic Laplacian $\widetilde{\nabla}^2$ are jointly parameterized by $s_1$, $s_2$, and $\mu$:
\begin{equation}
    \mathbf{g}(s_1, s_2) = \exp(\mathbf{S}), \qquad
    \mathbf{S} =
    \begin{pmatrix}
    s_1 & s_2 \\
    s_2 & -s_1
    \end{pmatrix}, \qquad
    \widetilde{\nabla}^2\mathbf{u}^{(t)} = 
    \sqrt{|\mu\mathbf{I}|}^{-1}\,\nabla \cdot 
    \bigl(\mathbf{g}^{-1}\sqrt{|\mu\mathbf{I}|}\,
    \mu^{-1/2}\mathbf{g}^{-1}\nabla\mathbf{u}^{(t)}\bigr).
    \label{eq:riemannian_metric}
\end{equation}
Intuitively, the axial stretching parameter $s_1$ controls the elongation ratio between the two local coordinate axes in calculating the local perception operation, steering the NCA to grow features that are directionally dominant along the horizontal or vertical direction. Similarly, the shearing parameter $s_2$ introduces diagonal distortion of the local coordinate system, controlling the dominant direction between the off-diagonal and diagonal orientations. The scale parameter $\mu$ uniformly changes the length scale of both local coordinate axes in the perception operation, controlling the size of morphological features. When $\mu = 1$, this Riemannian control reduces to pure anisotropy steering, controlled by $s_1$ and $s_2$ alone. When $s_1 = s_2 = 0$, the metric $\mathbf{g}$ becomes the identity, and the Riemannian control reduces to pure scale steering. The scale and anisotropy controls are therefore naturally unified in this Riemannian metric and can be applied together or separately.

\begin{figure}
  \centering
  \includegraphics[width=0.8\textwidth]{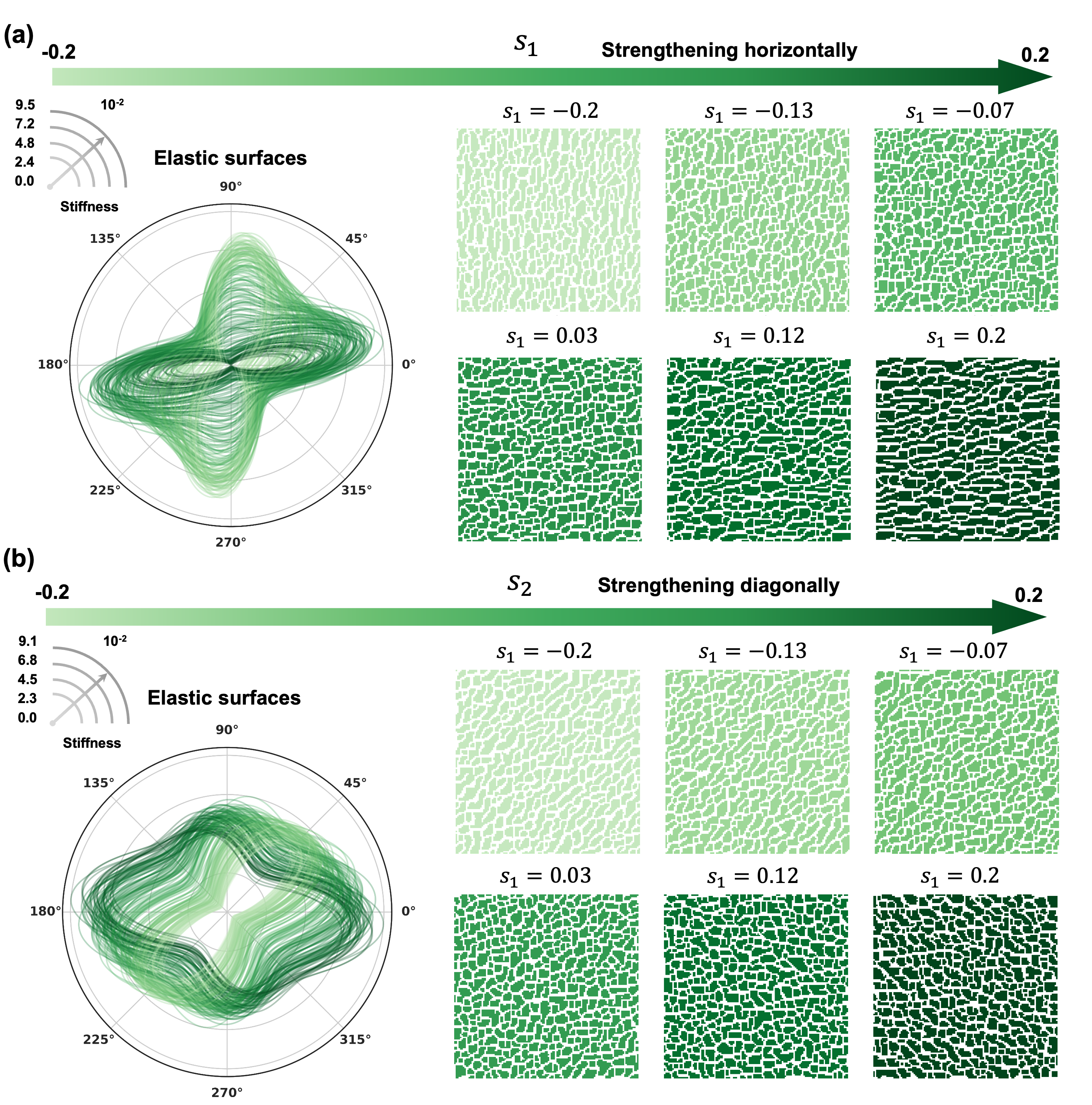}
  \caption{Generated microstructures and corresponding elastic surfaces under anisotropy steering, following the same layout as Fig.~\ref{fig_6}. \textbf{(a)} With increasing $s_1$, the principal stiffness shifts progressively from the vertical to the horizontal direction, demonstrating that $s_1$ provides dedicated control on anisotropy by manipulating the horizontal and vertical morphology. \textbf{(b)} With increasing $s_2$, the principal stiffness axes rotate between the off-diagonal and diagonal directions with a gradual change in the shear direction, demonstrating that $s_2$ provides an independent control over shear-dominated anisotropy that $s_1$ alone cannot access.}\label{fig_8}
\end{figure}

\begin{figure}
  \centering
  \includegraphics[width=0.8\textwidth]{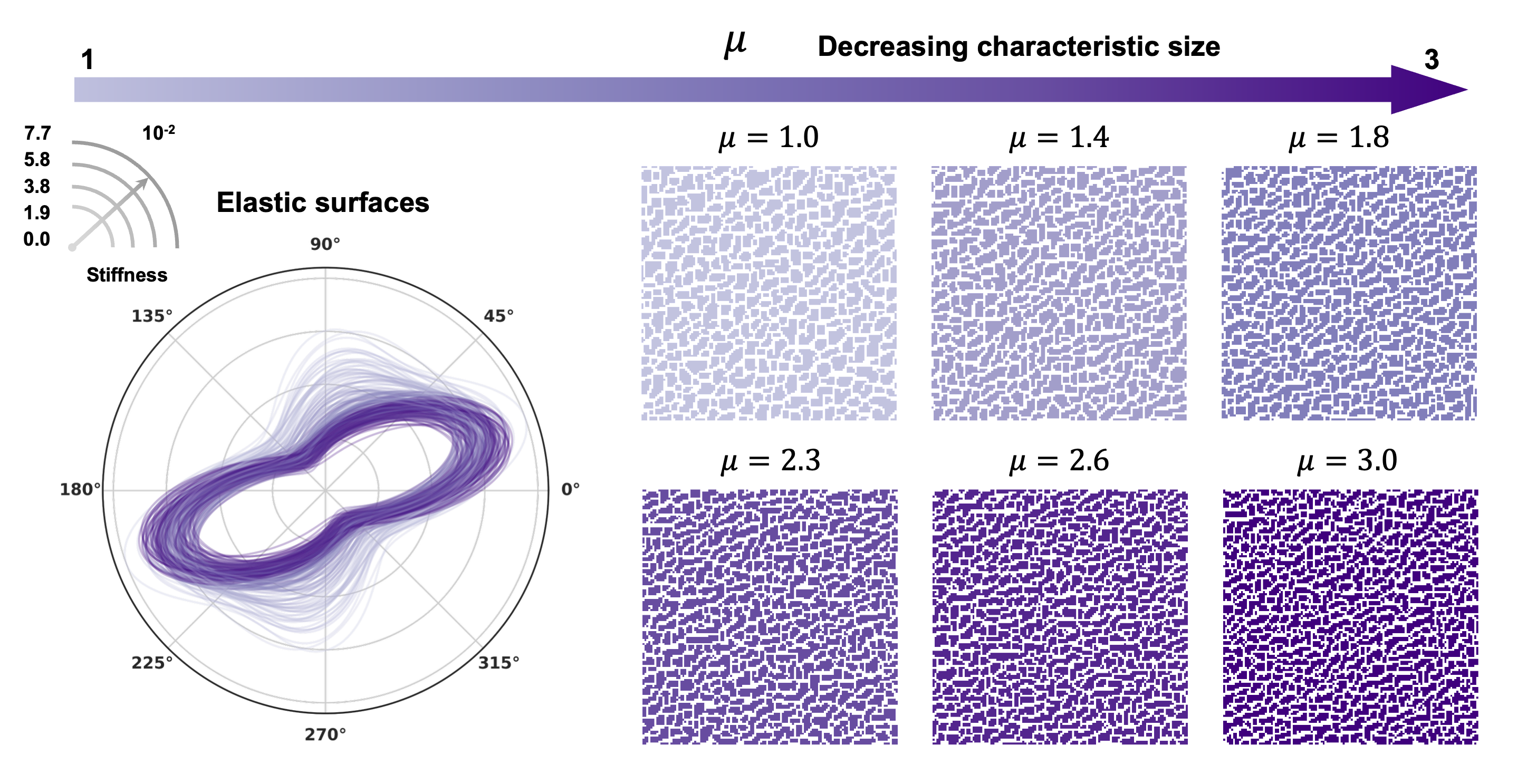}
  \caption{Generated microstructures and corresponding elastic surfaces under scale steering, following the same layout as Fig.~\ref{fig_6}. With increasing $\mu$, the characteristic feature size of the generated microstructure decreases progressively while the degree of mechanical anisotropy increases, as the inherent directional bias of the learned local rule accumulates more evidently at finer scales.}\label{fig_9}
\end{figure}

We use the leaf microstructures to illustrate this control mechanism. As shown in Fig.~\ref{fig_8}, $s_1$ and $s_2$ provide complementary control over axial and diagonal anisotropy of the generated microstructures, as intended. With increasing $s_1$, the generated microstructure continuously transitions from vertically dominated to horizontally dominated, and the principal direction of stiffness changes correspondingly from the vertical to the horizontal axis. As $s_2$ increases, the microstructure transitions from the off-diagonal to the diagonal direction, with the principal stiffness direction rotating accordingly. Together, $s_1$ and $s_2$ span the full directional anisotropy space through two control axes, producing continuous changes in both morphology and elastic response. Meanwhile, as shown in Fig.~\ref{fig_9}, increasing $\mu$ progressively refines the feature scale, while the degree of stiffness anisotropy and its magnitude remain largely unchanged.

\begin{figure}
  \centering
  \includegraphics[width=0.85\textwidth]{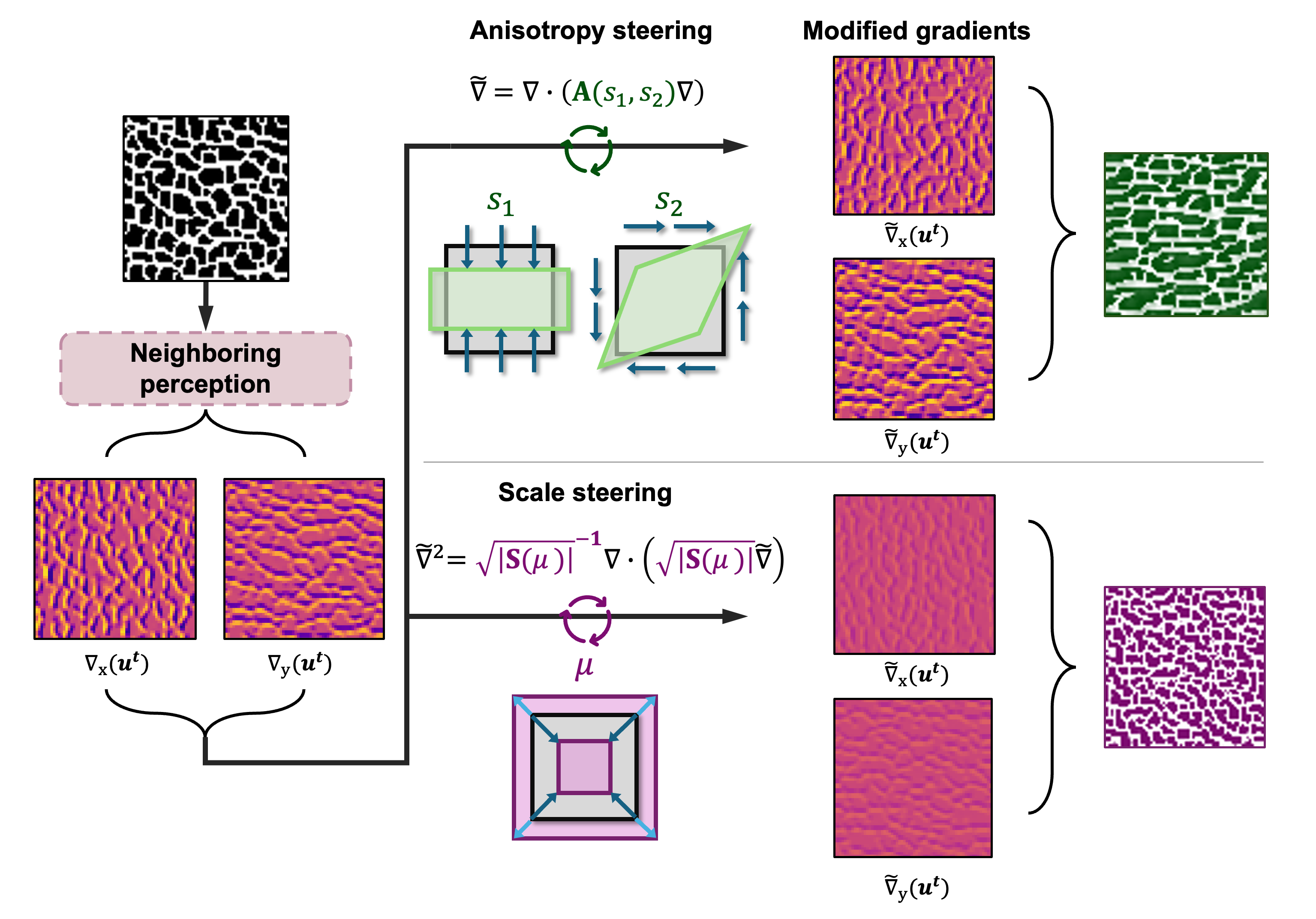}
  \caption{Anisotropy and scale controls applied to the NCA perception stage. The original perception is modified by a Riemannian metric tensor $\mathbf{g}(s_1, s_2)$ parameterized by the stretching parameter $s_1$ and shear parameter $s_2$, which reshapes the local perceptual space to stretch or shear the extracted gradient and Laplacian information before passing it to the trained update network. Scale steering additionally contracts the perceptual length scale uniformly through a scale factor $\mu$, making the perception stage extract neighborhood information as if the domain were enlarged or squeezed. Both mechanisms modify only the perception stage and can be applied simultaneously or independently without any retraining of the update network.}\label{fig_7}
\end{figure}

\subsubsection{Directional thickness control}
\begin{figure}
  \centering
  \includegraphics[width=0.85\textwidth]{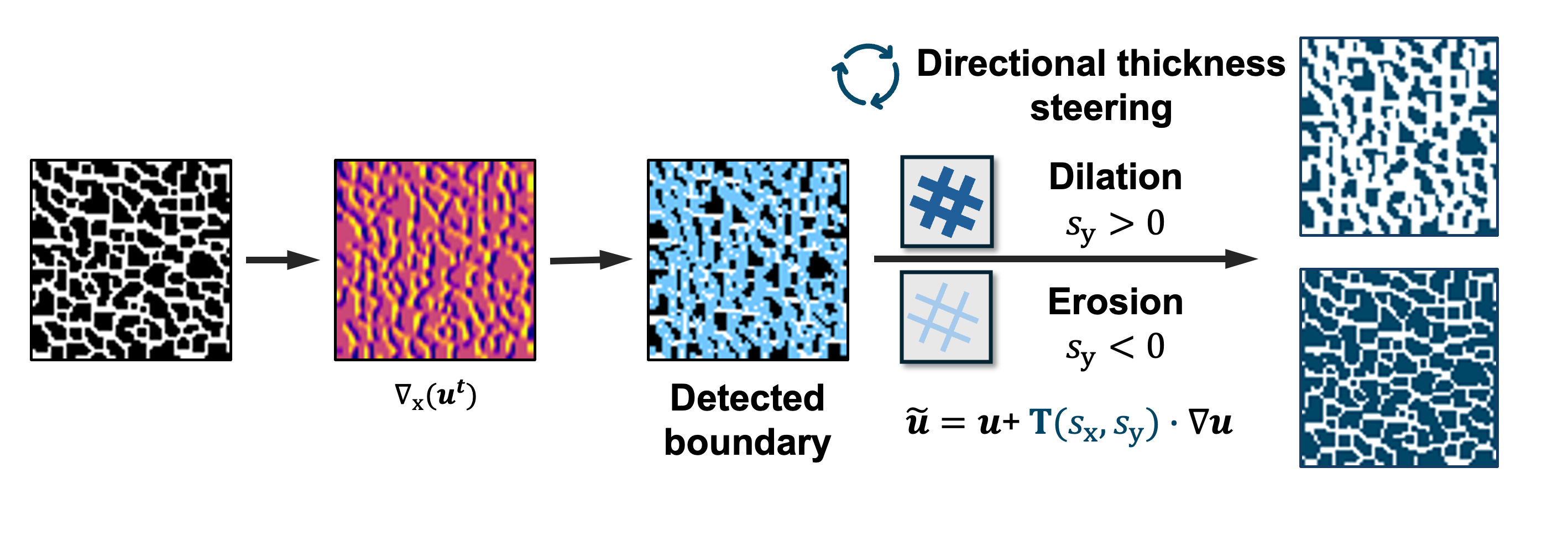}
  \caption{Directional thickness control applied as a post-processing boundary evolution stage. After the NCA generation, the solid-void boundary is detected through local gradients, using the same Sobel filters in the perception stage. A signed speed operator parameterized by the horizontal speed $s_x$ and vertical speed $s_y$ then drives the boundary outward to dilate or inward to erode the solid phase directionally, with positive values increasing and negative values decreasing the thickness along the corresponding axis. The evolution is confined strictly to the detected boundary regions and leaves the interior of each phase unaffected, preserving the overall network topology of the generated microstructure while providing independent control over the thickness in the horizontal and vertical directions, respectively.}\label{fig_10}
\end{figure}

The three steering mechanisms introduced above operate on the perception stage and control the morphological features of the generated microstructure, but do not directly regulate the solid volume fraction or feature thickness that governs the overall stiffness level. We therefore implement directional thickness control by making a few additional growth steps in the binarized solid-void field $\mathbf{u}$, as shown in Fig.~\ref{fig_10}. At each evolution step, the solid boundary is detected via the same Sobel filters as in the perception stage and updated as
\begin{equation}
    \tilde{\mathbf{u}} = \mathbf{u} + 
    \bigl(s_x \left| \nabla_y \mathbf{u} \right| + 
    s_y \left| \nabla_x \mathbf{u} \right|\bigr).
    \label{eq:grassfire_update}
\end{equation}
The absolute gradient magnitudes confine the update strictly to the solid-void interface, leaving the interior of each phase unaffected. And $s_x$ and $s_y$ are signed speed parameters controlling the boundary propagation in the horizontal and vertical directions, respectively. Positive $s_x$ or $s_y$ drives outward propagation of the solid boundary, dilating the solid phase and increasing the thickness, while negative values drive inward erosion. The two parameters act independently, enabling anisotropic control over the directional thickness of the solid phase. The evolution is iterated for a fixed number of steps and followed by adaptive binarization to recover a clean binary field.

\begin{figure}
  \centering
  \includegraphics[width=0.80\textwidth]{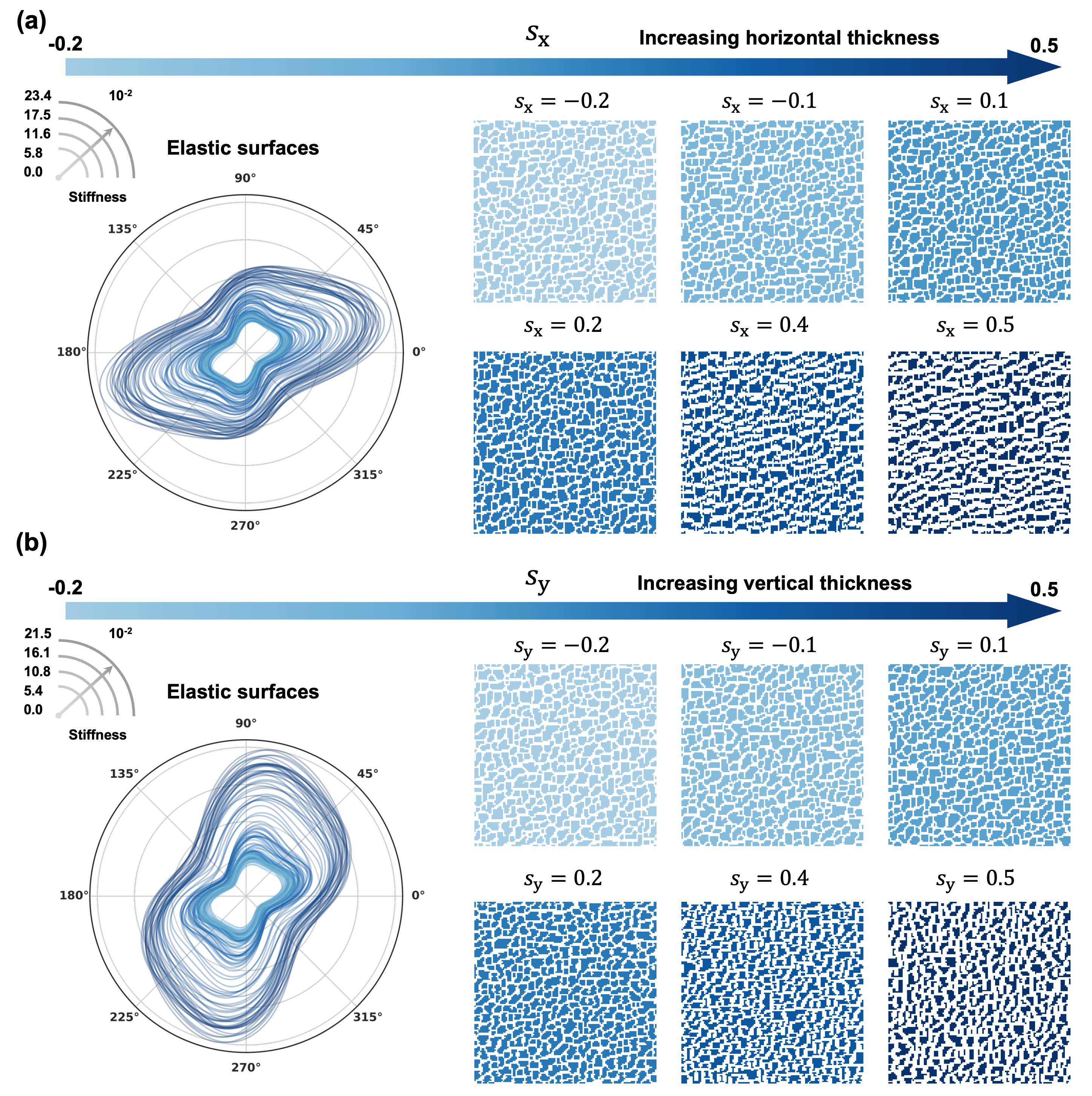}
  \caption{Generated microstructures and corresponding elastic surfaces under directional thickness steering, following the same layout as Fig.~\ref{fig_6}. \textbf{(a)} and \textbf{(b)} show the results for $s_x$ and $s_y$, respectively. Both parameters produce a consistent trend that increasing the control parameters thickens the solid struts, raises the overall stiffness level, and preferentially strengthens the stiffness along the corresponding dilation direction, with $s_x$ and $s_y$ acting independently along the horizontal and vertical axes respectively.}\label{fig_11}
\end{figure}

As shown in Fig.~\ref{fig_11}, both $s_x$ and $s_y$ produce two coupled effects on the mechanical properties: increasing the solid volume fraction raises the overall stiffness level, while the directional boundary evolution preferentially strengthens the stiffness along the controlled axis. The two parameters act independently along the horizontal and vertical directions, providing simultaneous control over the global stiffness level and the degree of mechanical anisotropy.

\subsection{Property space exploration with combined control}\label{subsubsection:design space}
In the previous sections, we have introduced four different post-training steering mechanisms, including orientation, anisotropy, scale, and directional thickness steering. Each of them can control the generated microstructures without retraining the NCA model. Here, we combined these mechanisms by randomly sampling control parameters $s_1$, $s_2$, $s_x$, and $s_y$ to explore the property space covered by their joint control, which lays the foundation for the multiscale design presented later.

Note that orientation and scale steering are intentionally excluded from the exploration. Orientation steering is excluded because it would introduce a rotational degree of freedom that couples with the directional effects of $s_1$, $s_2$, $s_x$, and $s_y$, making it difficult to isolate each parameter's contribution to the stiffness distribution. Moreover, as shown earlier, such orientational changes essentially amount to rotating the stiffness tensor. Scale steering is excluded because it has negligible effect on the mechanical properties. The generated dataset therefore spans a four-dimensional control space $[s_1, s_2, s_x, s_y]$, with each sample consisting of the control parameters, the generated microstructure, and its stiffness matrix $\mathbf{C}$.

\begin{figure}
  \centering
  \includegraphics[width=0.80\textwidth]{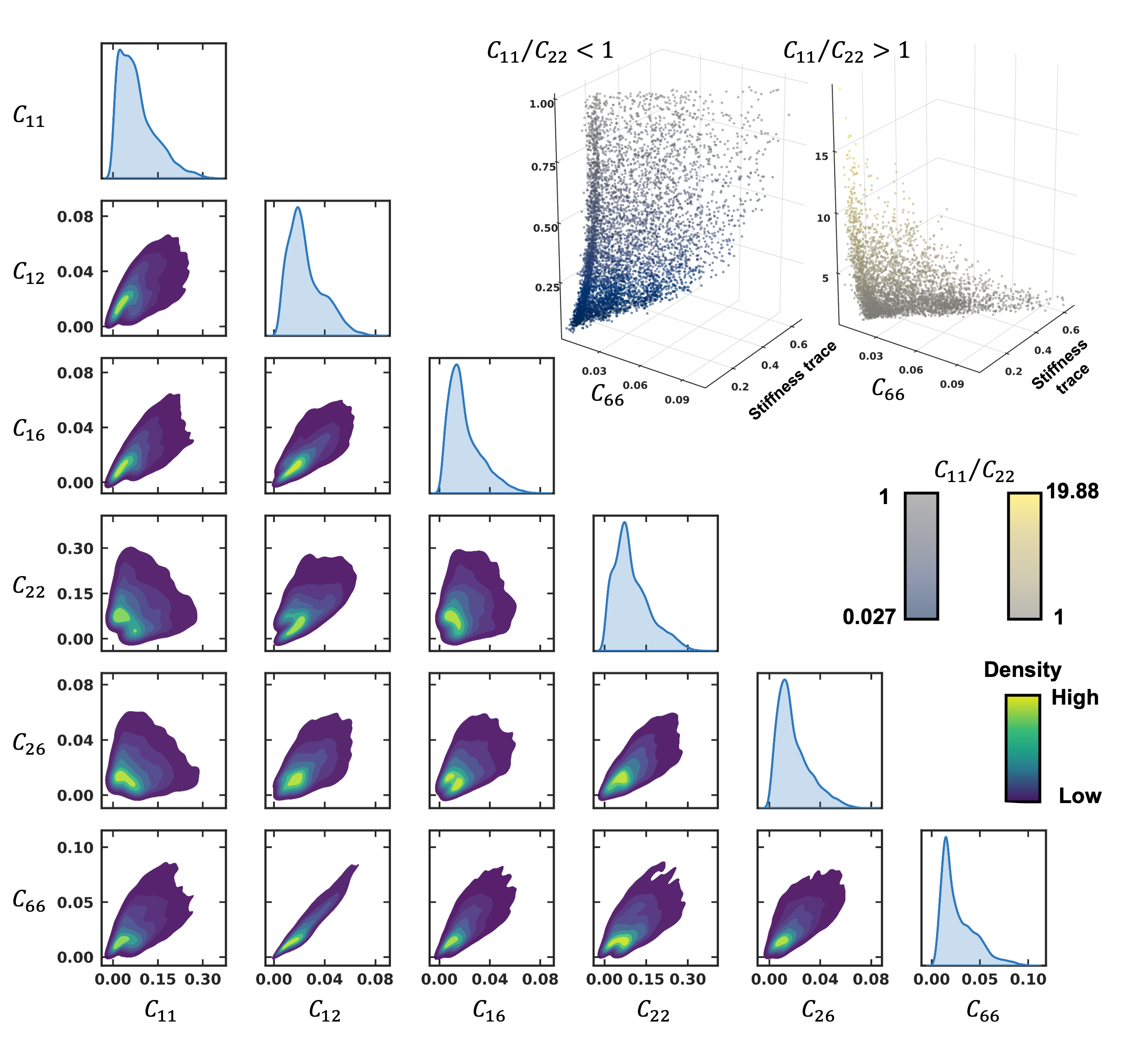}
  \caption{Mechanical property distribution of the generated dataset. The pairwise distribution matrix shows the joint and marginal distributions of the six independent stiffness components $C_{11}$, $C_{12}$, $C_{16}$, $C_{22}$, $C_{26}$, and $C_{66}$ across 10,000 generated microstructures, where the diagonal panels show the marginal distribution of each individual component and the off-diagonal panels show the joint distribution between each pair of components, respectively. The anisotropy of each sample is characterized by the ratio $C_{11}/C_{22}$, with samples colored by whether they are horizontally dominant ($C_{11}/C_{22} > 1$) or vertically dominant ($C_{11}/C_{22} < 1$), as illustrated in the three-dimensional scatter plots. The explored property space spans an anisotropy ratio ranging from $C_{11}/C_{22} = 0.027$ to $C_{11}/C_{22} = 19.88$, covering nearly three orders of magnitude in directional stiffness contrast, confirming that the combined steering controls provide access to a continuously organized, and mechanically diverse property space.}\label{fig_12}
\end{figure}

Fig.~\ref{fig_12} shows the pairwise and marginal distributions of the six unique stiffness components, $C_{11}$, $C_{12}$, $C_{16}$, $C_{22}$, $C_{26}$, and $C_{66}$, across the property space that spans a broad range with different levels of anisotropy. We use the ratio $C_{11}/C_{22}$ to quantify anisotropy,  where values above one indicate a horizontally dominant response and values below one a vertically dominant one. This ratio ranges from $0.027$ to $19.88$, a variation of nearly three orders of magnitude, further quantitatively validating the broad range of anisotropy achieved by the proposed control.

\begin{figure}
  \centering
  \includegraphics[width=0.85\textwidth]{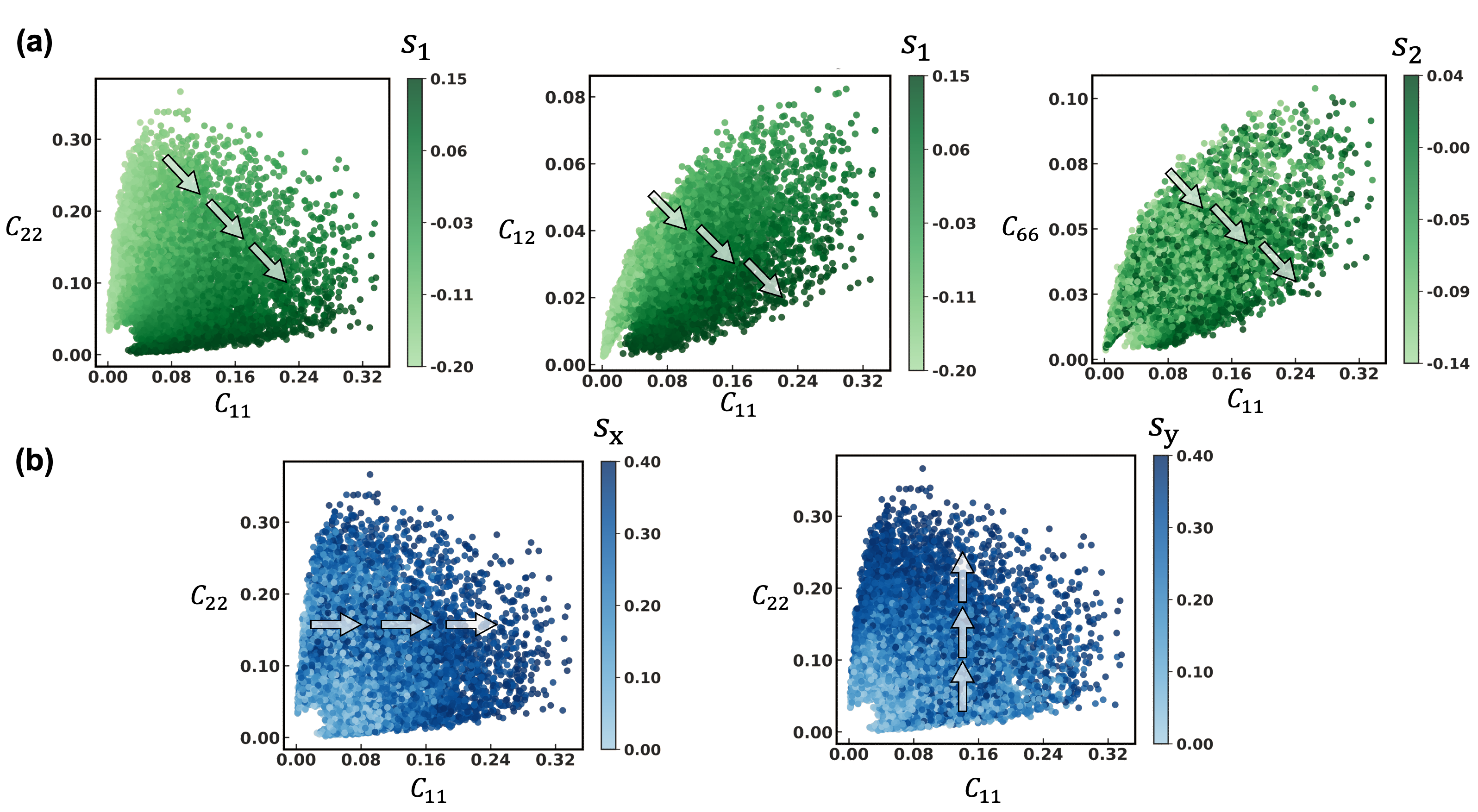}
  \caption{Individual controlling effect of each control parameter on the stiffness distribution. Each panel shows the distribution of a pair of stiffness components, colored by the value of the corresponding control parameter. \textbf{(a)} Increasing the stretching parameter $s_1$ shifts samples from vertically to horizontally dominant anisotropy, while increasing the shearing parameter $s_2$ independently governs the shear stiffness $C_{66}$, extending the reachable property space toward shear-dominated anisotropy in a direction that $s_1$ alone cannot access. \textbf{(b)} The directional thickness parameters $s_x$ and $s_y$ preferentially strengthen $C_{11}$ and $C_{22}$ along the horizontal and vertical axis, respectively. In all cases, the smooth transition across the distribution confirms that each control parameter acts continuously on the stiffness components.}\label{fig_13}
\end{figure}

The contribution of each individual parameter to the mechanical property space is illustrated in Fig.~\ref{fig_13}. The stretching parameter $s_1$ manipulates the anisotropy determined by $C_{11}$ and $C_{22}$, controlling whether the stiffness is horizontally or vertically dominant, while the shear parameter $s_2$ independently governs the shear stiffness $C_{66}$, extending the reachable property space toward shear-dominated stiffness that $s_1$ alone cannot access. The thickness parameters $s_x$ and $s_y$ produce complementary directional effects on the overall stiffness level, with $s_x$ preferentially raising $C_{11}$ and $s_y$ preferentially raising $C_{22}$. Together, the four parameters provide complementary and continuous access to a broad property space, as confirmed by the smooth color gradients observed across all parameter sweeps in Fig.~\ref{fig_13}. This organized property space forms the foundation for the multiscale design problem presented in Section~\ref{sec:multiscale}.

\section{Spatially varying growth and discretization transfer}\label{sec:extension}
In this section, we extend the generative framework beyond homogeneous microstructures and regular grids. We first show that, by using spatially varying control parameter fields, trained NCAs can grow heterogeneous microstructures with smooth transitions between morphologies and types. We then transfer the trained model from the regular grid on which it was trained to the complex 3D surfaces with irregular mesh by adapting the local perception, without retraining the model. Finally, we demonstrate the programmable growth process by user-defined sources and sequence, enabling growth in irregular geometry and in-situ repair of damaged microstructures.

\subsection{Spatially varying control and microstructure transition}\label{subsection:transition}
\begin{figure}
  \centering
  \includegraphics[width=0.85\textwidth]{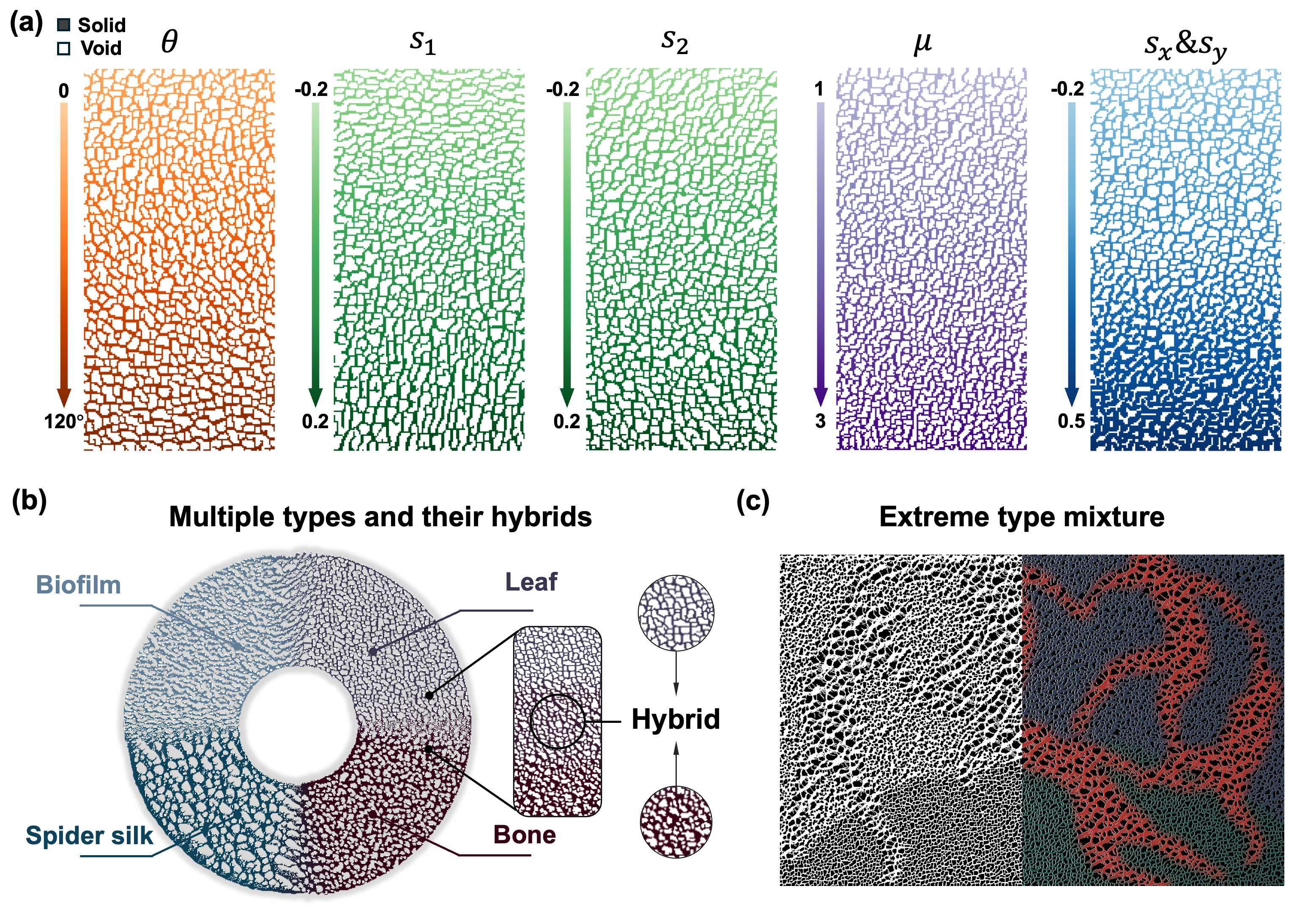}
  \caption{Spatially graded control and microstructure transition in the NCA framework. \textbf{(a)} Smooth transitions produced by spatially varying individual control parameters. Each parameter varies continuously from top to bottom across the domain, leading to gradual changes in orientation, anisotropy, scale, and directional thickness without sharp interfaces. \textbf{(b)} Smooth transitions between distinct microstructure types generated by blending independently trained NCA models during the iterative growth. The microstructure ring demonstrates continuous transitions among four distinct microstructure types. Hybrid microstructure prototype emerges in the transition region with merged morphological features from both parent types. \textbf{(c)} An extreme case demonstrating the smooth transition among irregularly mixed microstructure types. The left half displays the raw result and the right half is colored to distinguish the different microstructure types (red for spider silk, green for leaf, and blue for bone). }\label{fig_14}
\end{figure}

As demonstrated in Sec.~\ref{subsection:control}, uniform control parameters enable controlled growth of homogeneous microstructures with statistically equivalent morphology throughout the domain. Many applications, however, require heterogeneous microstructures with spatially varying morphological features to accommodate local requirements. To enable such designs, we extend the steering strategy by prescribing the control parameters locally, allowing them to vary spatially across the domain. This capability is a direct consequence of the local and self-organizing nature of the NCA. When the control parameters are prescribed as smoothly varying spatial fields, each cell follows its local steering conditions while remaining compatible with its neighbors through local information exchange. As a result, the generated morphology transitions smoothly across the domain without introducing discontinuities or requiring any post-processing. In Fig.~\ref{fig_14}(a), we first demonstrate such spatially varying growth for the individual control mechanisms. In each case, the corresponding control is prescribed as a smoothly varying spatial field from top to bottom. The same NCA then grows the microstructure under these local steering conditions, producing continuous morphological variations across the domain. 

Beyond spatially varying steering within one microstructure type, the framework can also generate transitions between distinct microstructure types. The ring-shaped example in Fig.~\ref{fig_14}(b) demonstrates smooth transitions among biofilm, leaf, bone, and spider silk microstructures. We achieve this by introducing a spatial interpolation mask during growth to blend the local update responses of multiple NCA models trained on different parent microstructure types. Because this blending is integrated into the iterative growth process rather than a direct interpolation of final generated structures, as in most existing methods, the smooth transition regions emerge naturally by design and inherit morphological features from both parent microstructures. In the transition region between leaf and bone, for example, the generated morphology combines leaf-like network junctions with bone-like porous features, forming a hybrid microstructure prototype that does not exist in either parent class. This could enable new design possibilities beyond naturally occurring or individually learned microstructures.

To stress test this transition capability, we further consider an extreme case with irregularly mixed microstructure types, as shown in Fig.~\ref{fig_14}(c) and Vid.~\ref{vid:type transition}. The result shows that the model remains stable under such a complex spatial layout with a smooth transition. Together, these demonstrations extend the framework from global microstructure generation to spatially programmed heterogeneous design.

\subsection{Transfer across geometric domains and discretizations}
\begin{figure}
  \centering
  \includegraphics[width=0.9\textwidth]{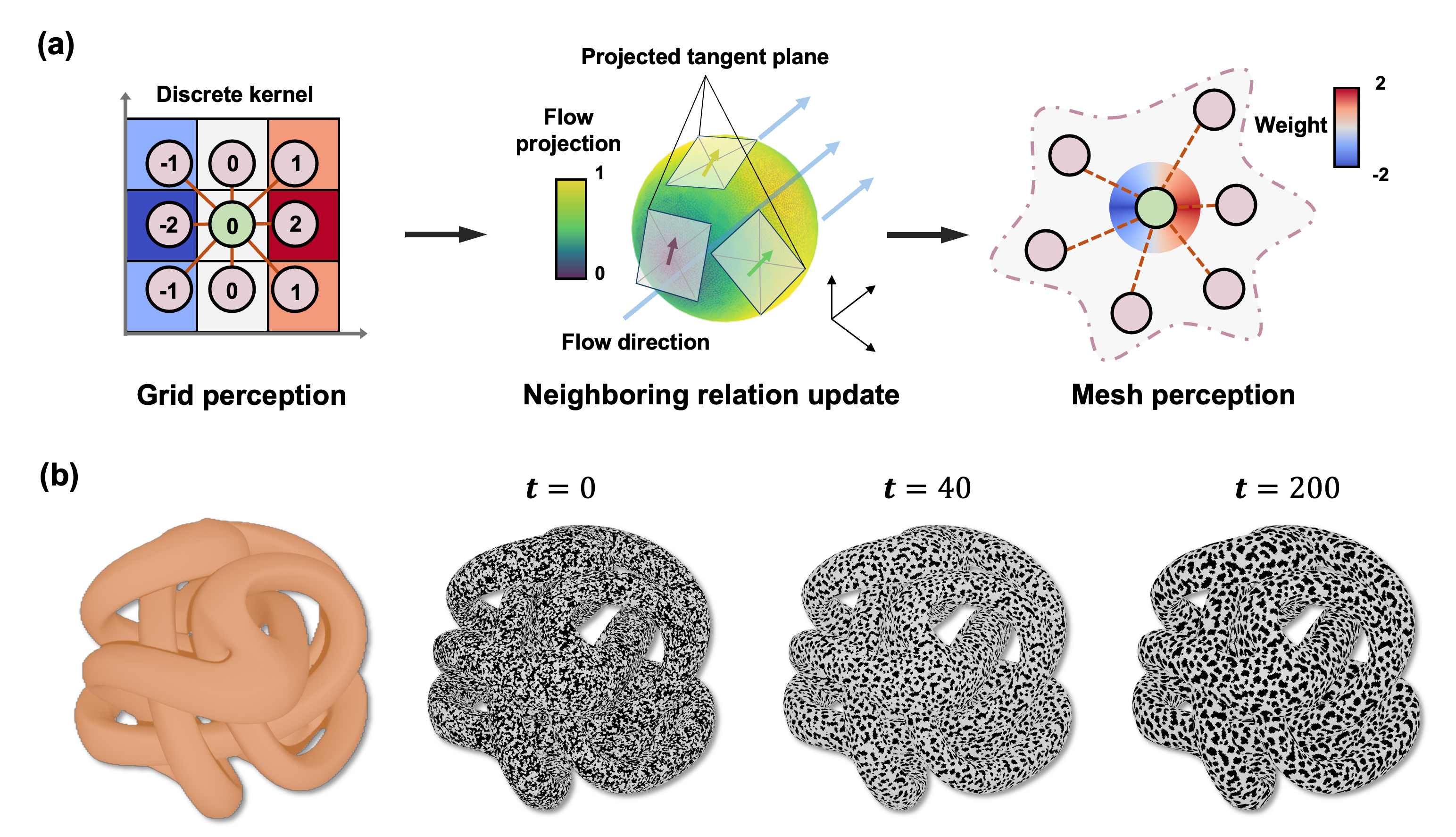}
  \caption{Transfer of the trained NCA framework from a uniform grid discretization to a surface-mesh discretization by redefining local perception operators. \textbf{(a)} A prescribed global flow direction $\mathbf{d}$ is projected onto the local tangent plane at each mesh vertex to define a consistent local coordinate frame. The fixed discrete kernels used on the uniform grid are replaced by continuous angular weights computed from the local frame with projected neighboring vertices. The horizontal gradient channel $\nabla_x$ is shown as an example, illustrating the correspondence between discrete grid weights and continuously varying mesh weights. \textbf{(b)} Growth on a complex 3D surface mesh. Starting from a random initial state, the model evolves through iterative local updates and grows a consistent microstructure that follows the surface geometry while maintaining the microstructure morphology.}\label{fig_15}
\end{figure}

All previous cases assume growth on a regular pixelated grid. We extend the framework to grow on irregular geometric domains with irregular discretization as shown in Fig.~\ref{fig_15}, broadening its applicability across design cases. For example, implant design often requires the design to adapt to irregular, patient-specific bone shapes~\cite{McAnena2025}. The key idea of such an extension builds on the PDE interpretation of the NCA dynamics we introduced in Sec.~\ref{subsubsection:NCA models}. We note that the same PDE can be solved and converge to the same result regardless of the domain or discretization, as long as the local operators are adapted. For example, a PDE yields the same solution whether solved on a triangular mesh, a hexagonal mesh, or even a point cloud, provided the local operators are redefined in the corresponding form that makes the solution compatible across domains. The governing PDE itself does not change, and only the local operators adapt to different instantiations. We generalize this idea to the NCA. NCA uses a neural network to map a set of local perception operators to the update. The neural network can thus be regarded as the universal PDE form. And to adapt the NCA to a new domain or discretization, we simply adapt the local perception operators, without retraining of the neural network. We take the irregular 3D surface mesh as an example to illustrate the transfer procedure.

Transferring the trained NCA to an irregular surface mesh faces two challenges that do not arise on a regular grid. First, local perception operators such as the $x$-gradient are directional, which requires a well-defined coordinate system to provide direction information. Unlike a consistent global coordinate system that generalizes for the whole grid, there is no such global system on the curved surface mesh because of the lack of global axis alignment. To address this issue, a consistent local frame must be established at every vertex so that the directions carry the same meaning across the surface. Second, the fixed weights of the grid kernels are no longer valid on an irregular mesh. Each vertex has a different number of neighbors at different relative positions, so the weights of local perception operators must be redefined intrinsically from the local mesh geometry. We address these challenges following well-established differential geometry tools to redefine local perception operators in local frames at each vertex~\cite{knoppel2013globally, yang2020pfcnn, masci2015geodesic}.

To establish the local frame, we prescribe a global flow direction $\mathbf{d} \in \mathbb{R}^3$ that defines a reference orientation across the surface~\cite{knoppel2013globally, yang2020pfcnn}, as shown in Fig.~\ref{fig_15}(a). At each vertex $i$ with position $\mathbf{v}_i$ and surface normal $\mathbf{n}_i$, we project $\mathbf{d}$ onto the local tangent plane to obtain the local $x$-axis:
\begin{equation}
    \mathbf{f}_i = \frac{(\mathbf{n}_i \times \mathbf{d}) \times \mathbf{n}_i}{\|(\mathbf{n}_i \times \mathbf{d}) \times \mathbf{n}_i\|},
    \label{eq:tangent_flow}
\end{equation}
and define the local $y$-axis as $\mathbf{t}_i = \mathbf{n}_i \times \mathbf{f}_i$. Together, $(\mathbf{f}_i, \mathbf{t}_i, \mathbf{n}_i)$ form an orthonormal local frame at the vertex $i$, providing fixed local coordinate systems for directional perception operators across the surface.

With the local coordinate system of each vertex, we define an intrinsic, angularly-weighted stencil that generalizes the applied kernels to the local tangent frame~\cite{masci2015geodesic}. Each neighbor $j$ of the vertex $i$, with position $\mathbf{v}_j$, is projected into this local tangent frame to obtain its local 2D coordinates $(u_{ij}, v_{ij})$ relative to the vertex $i$:
\begin{equation}
    u_{ij} = (\mathbf{v}_j - \mathbf{v}_i) \cdot \mathbf{f}_i, \qquad
    v_{ij} = (\mathbf{v}_j - \mathbf{v}_i) \cdot \mathbf{t}_i,
    \label{eq:neighbor_projection}
\end{equation}
where $u_{ij}$ and $v_{ij}$ are the local $x$- and $y$-offsets of neighbor $j$ from the center vertex $i$. On the uniform grid, the Sobel and Laplacian filters assign fixed weights according to the axial and diagonal positions of neighboring pixels. On the mesh, we replace these fixed weights with continuous weights computed from the projected neighbor offsets:
\begin{equation}
    w^{x}_{ij} = \frac{2\,u_{ij}}{|u_{ij}| + |v_{ij}|}, \qquad
    w^{y}_{ij} = \frac{2\,v_{ij}}{|u_{ij}| + |v_{ij}|}, \qquad
    w^{\mathrm{lap}}_{ij} = \frac{2}{|u_{ij}| + |v_{ij}|},
    \label{eq:mesh_weights}
\end{equation}
and assign the center vertex the negative sum of its neighbor weights, ensuring that the operators vanish on constant fields. These intrinsic weights produce mesh analogs of the grid gradient and Laplacian kernels, and degenerate exactly to the 2D grid kernels, making the operators compatible with both discretizations. The resulting perception tensor is arranged in the same channel order as the grid perception, yielding a tensor of shape $[B, 4C, V]$ that is passed directly to the trained update network.


As shown in Fig.~\ref{fig_15}(b), we demonstrate the transferred growth on a complex 3D surface. Starting from a random initial state, the target morphology begins to emerge coherently and finally reaches a stable state. This result shows that the learned local update rule is not restricted to the uniform grid discretization on which it was trained, provided that local perception supplies compatible local differential information on the target discretization.

\subsection{Programmable asynchronous growth}\label{subsection:asynchronous}
Beyond controlling the morphology itself, we introduce an additional level of control to NCA by prescribing a growth sequence to control where and when growth occurs. Within the activated region, the NCA dynamics evolve as usual, while cells outside remain frozen. When those frozen regions are activated later, the self-organizing dynamics allow them to interact with the previously  activated regions, merging multiple evolutions into one to ensure compatible connections and achieve asynchronous growth. This mirrors developmental processes in nature, where tissues grow at different times and rates yet still merge seamlessly, much as a healing wound regenerates and reintegrates with the surrounding tissue. We demonstrate this capability through two representative examples.

\subsubsection{Two-source growth in complex geometry}
\begin{figure}
  \centering
  \includegraphics[width=1.0\textwidth]{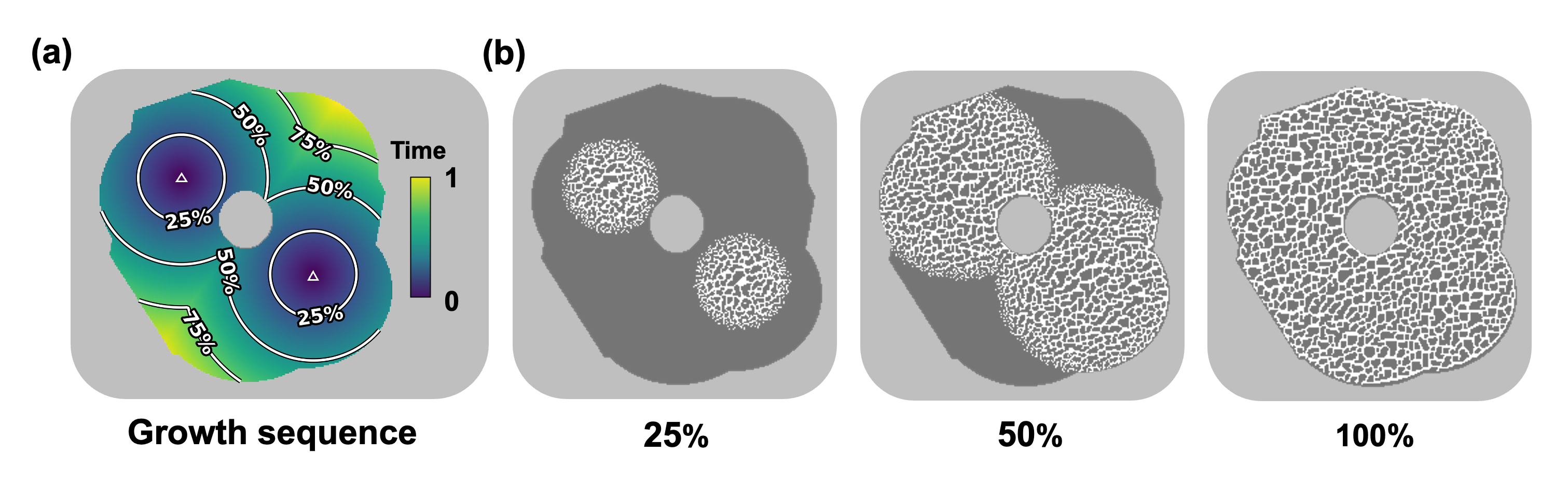}
  \caption{Asynchronous growth in an irregular domain guided by prescribed growth sources and sequence. \textbf{(a)} Two source points (triangle marks) define the growth sequence that activates cells progressively from the sources outward. Once the cells are activated, they keep growing and interacting with their neighboring cells. \textbf{(b)} The growth process shows the two activation fronts expanding with time, merging where their influence regions meet, eventually covering the full irregular domain, growing consistent microstructures.}\label{fig_16}
\end{figure}

The first example demonstrates asynchronous growth inside an irregular domain, as shown in Fig.~\ref{fig_16}. We represent the domain by a binary mask $\mathbf{M} \in \{0,1\}$, where cells outside the mask remain fixed at zero throughout growth. We then prescribe two source points, $\mathbf{m}_1$ and $\mathbf{m}_2$, inside the domain to define where growth begins and how the active region expands. The growth sequence is determined by a normalized nearest-source distance field:
\begin{equation}
    Dist(\mathbf{p}) = \frac{\min_k \|\mathbf{p} - \mathbf{m}_k\|_2}{\max_{\mathbf{p}' \in \mathbf{M}} \min_k \|\mathbf{p}' - \mathbf{m}_k\|_2},
    \label{eq:distance_field}
\end{equation}
where cells closer to either source have lower values of $Dist(\mathbf{p})$ and start to grow earlier. The sequence allows the growth to expand outward from both sources simultaneously, as shown in Fig.~\ref{fig_16}(a). The source(s) and growth sequence can always be customized for each specific case as required, serving as a flexible interface for programmable growth. 

As shown in Fig.~\ref{fig_16}, the two growth fronts expand from the prescribed sources, merge where their influence regions meet, and eventually fill the irregular domain. We illustrate the whole growth process in Vid.~\ref{vid:asynchronous growth}. The result remains coherent and consistent while following the prescribed growth sequence. This example shows that customized growth sources and sequence can program the growth spatially and temporally  in arbitrary domain shapes. The asynchronous growth allows parallel generation starting from multiple sources in large-scale structures, without retraining or modifying the learned NCA update network. By enabling only the active portion to grow, i.e., the growth front of each source, it reduces peak memory consumption and enables straightforward parallelization. In contrast to generating the entire domain all at once, which may not be scalable, this provides an alternative path that helps the generation scale to large and complex domains.

\subsubsection{In-situ microstructure repair}
\begin{figure}
  \centering
  \includegraphics[width=0.9\textwidth]{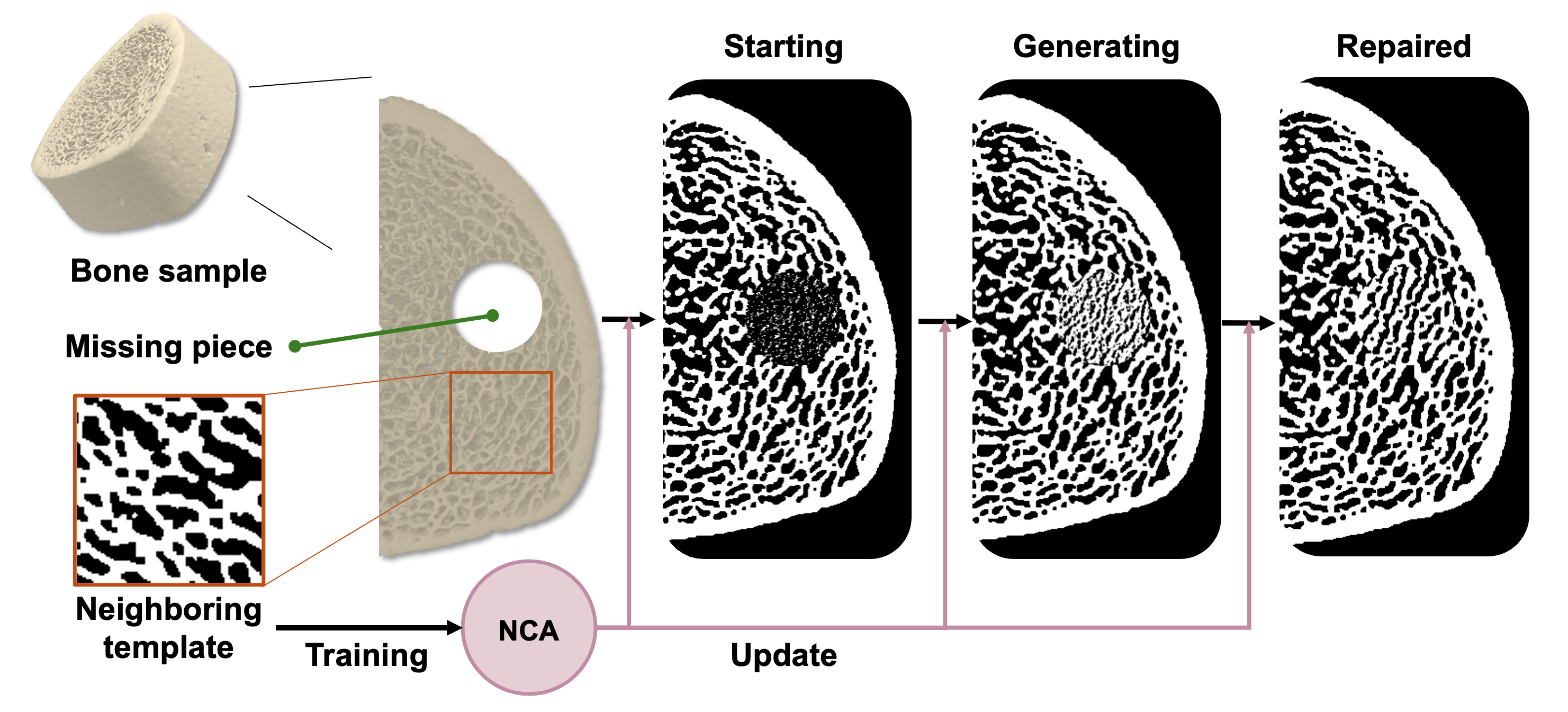}
  \caption{In-situ repair of a damaged trabecular bone microstructure utilizing the NCA generation framework. A circular region is removed from a real cross-sectional bone slice to simulate a localized structural defect. An NCA model is trained on the neighboring region surrounding the defect as the template, learning the local morphology of the surrounding bone microstructure. The repair proceeds by restricting the asynchronous growth to the missing region while the surrounding intact area remains fixed as the boundary condition, allowing the NCA to grow an infill that is seamlessly integrated with the surrounding microstructure at the repair boundary.}\label{fig_17}
\end{figure}

Beyond controlling the sequence in irregular domains, asynchronous growth can also be used to regrow a prescribed region surrounded by preexisting microstructures, enabling the in-situ repair of a missing region in a biological microstructure, as shown in Fig.~\ref{fig_17}. This is achieved by restricting growth to the missing region while keeping the surrounding intact area fixed to serve as the boundary condition of growth. The NCA is thus directed to grow only within the missing region, respecting the existing material boundaries. We demonstrate this on a cross-sectional slice of trabecular bone, from which a circular region is removed to simulate a localized structural defect. Rather than requiring a reference structure in the missing region, we extract a small adjacent intact patch as the single training template for NCA used for the repair. 

During repair, the NCA applies the stochastic asynchronous update in the missing region while the exterior remains fixed. The fixed region acts as a boundary condition, allowing the generated structure inside the defect to interact with and connect to the existing trabecular network at the repair interface. We illustrate the two-stage process from training to repairing in Vid.~\ref{vid:bone repair}. As a result, the generated microstructure inherits the local morphology of the neighboring template and embeds smoothly with the surrounding bone. This example highlights two key advantages of the framework. First, the one-shot training feature enables the model to learn from a single small neighboring patch, without requiring large patches or any reference structure in the missing region. Second, asynchronous growth allows the new material to grow into the defect and connect to the existing trabecular network rather than forming a separate, mismatched patch. Together, these properties give the framework strong potential for patient-specific implant design, and can serve as a robust, self-restoring representation for other engineering systems where resilience against damage or attack is critical.

\section{Multiscale optimization}\label{sec:multiscale}
In this section, we integrate the proposed NCA generative framework into a multiscale optimization pipeline. We demonstrate this pipeline through a mechanical cloaking design, which features irregular domain geometries and complex heterogeneous property requirements. By following the growth cues provided by the optimization process, the NCA model steers its self-organizing dynamics to generate a heterogeneous multiscale structure with smooth transitions, accommodating complex geometries, and growing naturally across different resolutions without requiring a redesign. This capability enables high-performance cloaking designs without the complicated post-processing or scalability limitations of existing methods, demonstrating the generality and scalability of the proposed framework.

\subsection{Optimization framework and problem setup}
\subsubsection{General optimization problem}
To translate the controllable microstructure generator into a structural design tool, we couple the NCA model with multiscale optimization. The goal is to generate spatially varying microstructures to achieve a target structural response or maximize the system-level performance. We discretize the macrostructure into a finite-element mesh of $n_e$ four-node quadrilateral elements. Each design element $e$ is assigned an effective stiffness matrix $\mathbf{C}_e$, constrained within the explored property space of a trained NCA in Sec.~\ref{subsubsection:design space} enabled by the proposed hybrid control strategy. The macroscale response is governed by
\begin{equation}
    \mathbf{K}(\mathbf{C}_e)\,\mathbf{U} = \mathbf{F},
    \label{eq:fe_equilibrium}
\end{equation}
where $\mathbf{K}$ is the global stiffness matrix assembled from the constitutive matrix in each element, $\mathbf{U}$ is the nodal displacement vector, and $\mathbf{F}$ is the applied load vector. The general multiscale optimization problem is defined as:
\begin{equation}
    \min_{\mathbf{C}_e} \quad f\!\left(\mathbf{U}(\mathbf{C}_e)\right)
    \label{eq:general_opt}
\end{equation}
\begin{equation*}
    \text{s.t.} \quad \mathbf{K}(\mathbf{C}_e)\,\mathbf{U} = \mathbf{F},
\end{equation*}
\begin{equation*}
    \mathbf{C}_e \in \mathcal{M}, \quad e \in \Omega_d,
\end{equation*}
where $\Omega_d$ is the set of design elements, $f(\cdot)$ is a given objective, and $\mathcal{M}$ denotes the explored property space. The elemental stiffness matrix $\mathbf{C}_e$ in the 2D setting is fully characterized by six independent stiffness components in Voigt notation $[C_{11}, C_{12}, C_{22}, C_{16}, C_{26}, C_{66}]$. 

For a given microstructure template, we can obtain the reachable property space of the trained NCA by varying the control parameters, as shown in Fig.~\ref{fig_12}. A straightforward approach is to directly optimize the stiffness of each element within this explored property space. However, while feasible, this strategy is not ideal and can be challenging in practice. The reason is that the feasible property space is highly skewed, as shown in Fig.~\ref{fig_12}, resulting in a high-dimensional irregular boundary. This makes it difficult to formulate effective constraints on the stiffness components. Even with such constraints, the optimizer may frequently violate the feasible region or become trapped during optimization. 

\begin{figure}%
  \centering
  \includegraphics[width=1\textwidth]{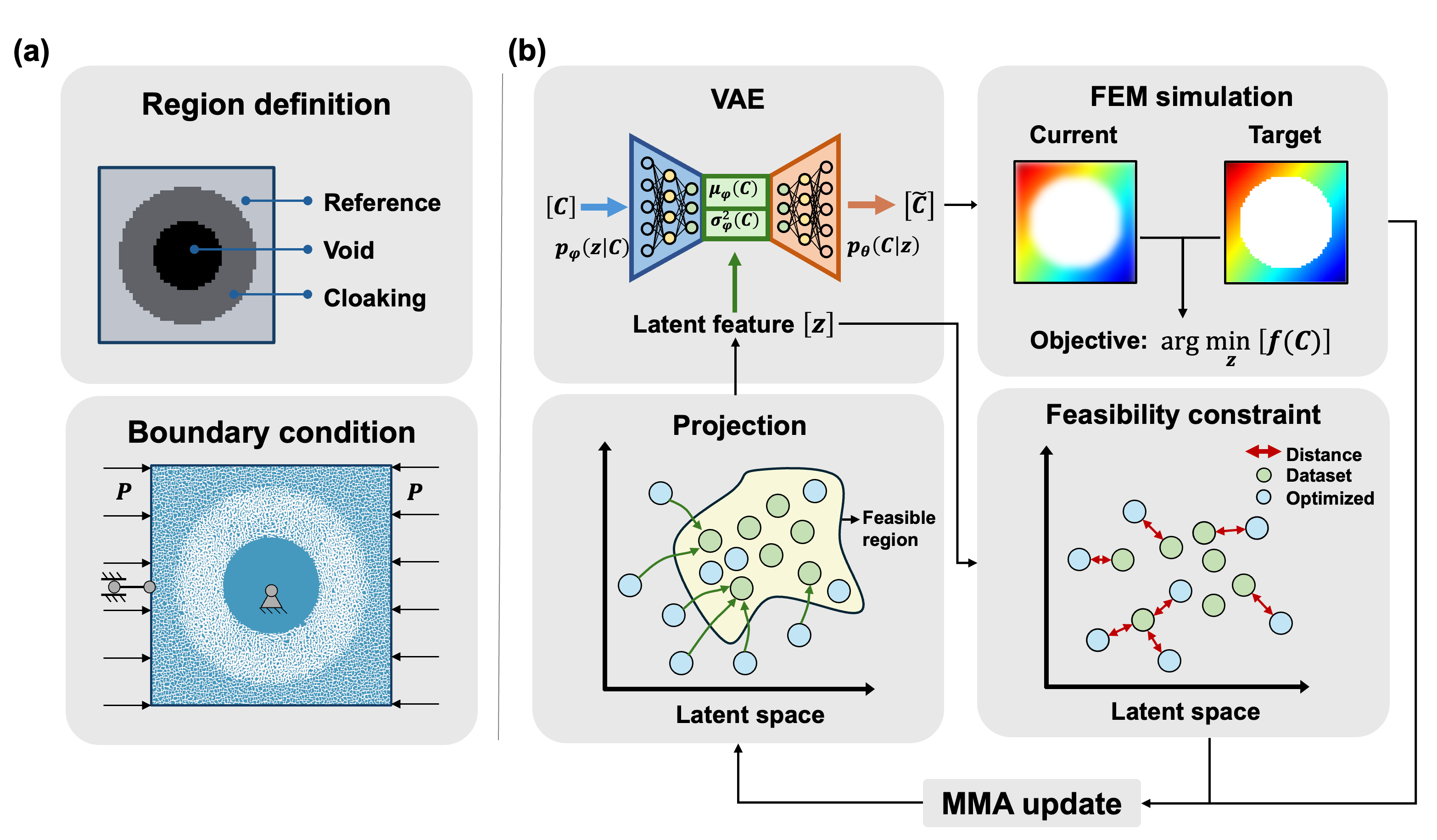}
  \caption{Optimization pipeline and problem setup for the mechanical cloaking case. \textbf{(a)} The macroscopic domain is divided into a cloaking region $\Omega_c$, a reference region $\Omega_r$, and a void region $\Omega_v$, loaded by compressive pressure on the left and right edges while rigid-body motion is suppressed by fixing the center node and restricting the vertical displacement of the left edge on its midpoint node. \textbf{(b)} At each optimization iteration, the latent field is decoded into stiffness for finite element simulation, and the adjoint sensitivity is chained through the VAE decoder Jacobian to update the latent variables via MMA. A latent-tube feasibility constraint keeps the optimized latent variables close to the explored property space, with out-of-tube cells projected back to their nearest samples. The loop iterates until the displacement field of the cloaked structure matches that of the homogeneous displacement fields over the surrounding reference region.}\label{fig_18}
\end{figure}

To address this issue, we introduce a variational autoencoder (VAE)~\cite{WangGenerative2020, Xiang2025} as a dimensionality reduction and regularization tool to construct a bidirectional mapping between the stiffness components and a lower-dimensional latent space $\mathbf{z}_e \in \mathbb{R}^{d_z}$. The neural network encoder maps the stiffness matrix $\mathbf{C}_e$ to the latent representation $\mathbf{z}_e$, while the paired decoder reconstructs the stiffness matrix from the latent variables, i.e. $\mathbf{C}_e = D(\mathbf{z}_e)$. Instead of optimizing and constraining the original stiffness components $\mathbf{C}_e$, we can now reformulate the optimization problem in the latent space, which is lower dimensional and exhibits a more regular distribution (ideally Gaussian). This transformation greatly simplifies constraint construction and improves optimization efficiency. Specifically, the optimization is reformulated in terms of the latent variables $\mathbf{z} = \{\mathbf{z}_e\}_{e \in \Omega_d}$:
\begin{equation}
    \min_{\mathbf{z}} \quad f\!\left(\mathbf{U}(\mathbf{z})\right)
    \label{eq:latent_opt}
\end{equation}
\begin{equation*}
    \text{s.t.} \quad \mathbf{K}(\mathbf{z})\,\mathbf{U} = \mathbf{F},
\end{equation*}
\begin{equation*}
    \mathbf g(z)\leq0,
\end{equation*}

\noindent where $\mathbf{g}(\mathbf{z})$ is a feasibility constraint in the latent space to ensure that the optimized design remains within the explored property space. Specifically, the feasible properties obtained in Sec.~\ref{subsubsection:design space} by varying the control parameters serve as anchor points that span the explored property space. We require the optimized latent variables to remain close to these anchors and formulate $\mathbf{g}(\mathbf{z})$ as a latent tube constraint. For each design element, we compute the normalized distance $\mathrm{dist}_e$ to the nearest feasible latent anchor and define the tube radius as $r_0 = \sqrt{d_z} \cdot r_{\mathrm{frac}}$, where $r_{\mathrm{frac}}$ is a given fractional value to decide the radius of feasible range. A smooth activation function is then used to penalize latent variables outside the tube:
\begin{equation}
    h_e = \frac{1}{2}\left(\tanh\!\left(\beta \left(\mathrm{dist}_e - r_0\right)\right) + 1\right),
    \label{eq:tube_activation}
\end{equation}
where $\beta$ increases progressively during optimization to sharpen the constraint boundary. The activation $h_e$ varies smoothly between 0 and 1: it approaches 0 when $\mathrm{dist}_e < r_0$, indicating that the latent variable lies safely inside the tube, and approaches 1 when $\mathrm{dist}_e > r_0$, indicating that it has moved outside the feasible neighborhood. Therefore, an ideal design requires $h_e$ to remain close to 0 for all design elements. Aggregating over the domain, the 
feasibility constraint is expressed as
\begin{equation}
    g(\mathbf{z}) = \frac{1}{|\Omega_d|}\sum_{e \in \Omega_d} h_e - \gamma \leq 0,
    \label{eq:tube_constraint}
\end{equation}
where $\gamma$ is the allowable fraction of elements outside the tube. In addition to this smooth constraint, as a further measure to ensure feasible designs, any element with latent variables that falls outside the tube after each optimization iteration is projected to its nearest feasible anchor points. We emphasize that the VAE is applied to the explored property space rather than the microstructure geometry. Its purpose is solely to facilitate the formulation of constraints in multiscale optimization. The microstructure geometry is learned by the NCA framework and still requires only a single template as the training data. 

The sensitivity of the objective with respect to the latent variables is obtained by chaining the adjoint sensitivity of $f$ with respect to the stiffness components through the Jacobian of the decoder:
\begin{equation}
    \frac{\partial f}{\partial \mathbf{z}_e}
    = \mathbf{J}_e^{\top}
    \frac{\partial f}{\partial \mathbf{C}_e}, \qquad
    \mathbf{J}_e = \frac{\partial D(\mathbf{z}_e)}{\partial \mathbf{z}_e},
    \label{eq:chain_rule}
\end{equation}
where $\mathbf{J}_e \in \mathbb{R}^{6 \times d_z}$ is the Jacobian of the decoder at $\mathbf{z}_e$, obtained by automatic differentiation, and $\partial f / \partial \mathbf{C}_e$ is the adjoint sensitivity. The sensitivity of the aggregated constraint can be obtained in a similar way. The sensitivities of the objective function and constraints enable the use of gradient-based optimizers to navigate the latent space iteratively, such as the Method of Moving Asymptotes (MMA) adopted in this study.


\subsubsection{Case study: mechanical cloak design}
While the proposed optimization is general and can accommodate different multiscale design requirements, we choose to demonstrate it on a mechanical cloak design~\cite{Senhora2025}, due to its demanding requirement for spatially varying microstructures with anisotropic mechanical properties in different regions, making it an ideal stress test of the framework's ability to translate optimized property fields into a heterogeneous structure with a smooth transition. 

In this case, the macrostructure is a $50 \times 50$ finite-element grid divided into three regions. As shown in Fig.~\ref{fig_18}(a), we design the mechanical cloak region $\Omega_c$ around a void $\Omega_v$ within the structure. The goal is to make the displacement field $\mathbf{U}_{\mathrm{ref}}$ in the surrounding reference region $\Omega_r$ retain the values it would have in a homogeneous structure without the void. The void region is fixed to a near-zero stiffness, and the reference region is fixed to the stiffness of the reference microstructure $\mathbf{C}^{\mathrm{ref}}$. The properties of the cloaking region $\Omega_c$ are optimized to minimize the void-induced distortion of the displacement field. We define the objective as:
\begin{equation}
    f(\mathbf{U}) =
    \frac{\left\|\mathbf{\gamma} \odot \left(\mathbf{U} - \mathbf{U}_{\mathrm{ref}}\right)\right\|_2^2}
    {\left\|\mathbf{\gamma} \odot \mathbf{U}_{\mathrm{ref}}\right\|_2^2 + \varepsilon},
    \label{eq:cloaking_obj}
\end{equation}
where $\mathbf{\gamma}$ is a binary mask to distinguish the reference region and $\varepsilon$ is a regularization constant. We apply constant compressive pressure on the left and right edges to induce deformation, while rigid-body motion is suppressed, as shown in Fig.~\ref{fig_18}(a). All design elements are initialized with the latent vector of reference property $\mathbf{C}^{\mathrm{ref}}$. The sensitivity of the objective with respect to the stiffness of each design element is computed using the adjoint method, which avoids solving the forward sensitivity equation separately for every design variable. The adjoint variable $\boldsymbol{\lambda}$ is obtained from the global system $\mathbf{K}^{\top}\boldsymbol{\lambda} = \partial f / \partial \mathbf{U}$. Substituting this adjoint solution and the explicit form of $\partial f/\partial \mathbf{U}$ derived from the objective in Eq.~\eqref{eq:cloaking_obj}, the full sensitivity with respect to stiffness component $C_{e,ij}$ of design element $e$ is:
\begin{equation}
\frac{\partial f}{\partial C_{e,ij}}
=
-\left[
\frac{2\,\boldsymbol{\gamma}\odot(\mathbf{U}-\mathbf{U}_{\mathrm{ref}})}
{\left\|\boldsymbol{\gamma} \odot \mathbf{U}_{\mathrm{ref}}\right\|_2^2 + \varepsilon}
\right]^{\top}
\mathbf{K}^{-1}
\frac{\partial \mathbf{K}}{\partial C_{e,ij}}
\mathbf{U}.
\label{eq:sensitivity}
\end{equation}
The complete optimization loop is summarized in Fig.~\ref{fig_18}(b). In iteration $t$, the current latent field $\mathbf{z}^t$ is decoded into an element-wise stiffness matrix $\mathbf{C}^t$, which is assembled into the global stiffness matrix for the finite element simulation. The resulting displacement field $\mathbf{U}^t$ is compared with the homogeneous reference field. The adjoint solve provides the sensitivity of the objective with respect to the element stiffness components, and the decoder Jacobian maps this sensitivity to the latent variables. The MMA optimizer then updates the latent field using the objective gradient together with the feasibility constraint. After each update, latent points that leave the feasible region are projected back to their nearest samples in the dataset. The loop continues until convergence, progressively tailoring the stiffness distribution in $\Omega_c$ to minimize the displacement mismatch in the reference region. Detailed setups are shown in Table~\ref{tbl_a2}.

\subsection{Optimized results}
\begin{figure}%
  \centering
  \includegraphics[width=1\textwidth]{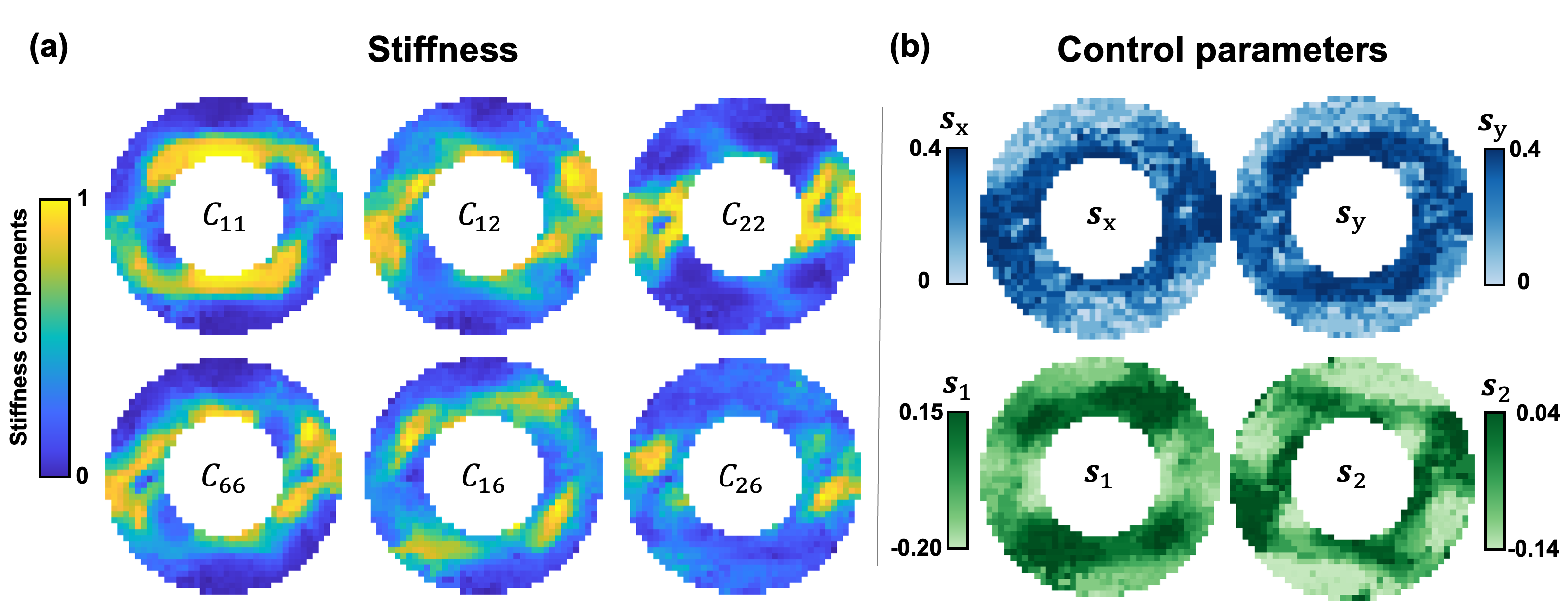}
  \caption{Optimized stiffness distribution and projected control parameter fields in the cloaking region. \textbf{(a)} Spatially varying distributions of the six stiffness components, where the angular distribution pattern reflects the directional stiffness redistribution required to redirect the stress field around the void. \textbf{(b)} Control parameter fields obtained by projecting the optimized latent variables to the nearest explored samples, which are passed directly to the trained NCA as the growth cues to produce the full cloaking structure.}\label{fig_19}
\end{figure}

The optimized stiffness component distributions and the corresponding NCA control parameter fields within the cloaking region are shown in Fig.~\ref{fig_19}(a). The stiffness components in Fig.~\ref{fig_19}(a) indicate a directional redistribution of the effective stiffness around the void, which manipulates the elastic response to minimize the void-induced distortion of the displacement field and recover the homogeneous displacement response in the reference region. Fig.~\ref{fig_19}(b) shows the NCA control parameter fields after projection onto the explored property space, where $s_1$ and $s_2$ control spatially varying anisotropy and $s_x$ and $s_y$ control directional thickness at each element. The projected control parameter fields will guide the growth of microstructures to realize the optimized stiffness distribution. In the homogenized simulation, it reduces the relative displacement error to $8.10\%$ compared with $129.41\%$ for the uncloaked structure, as reported in Table~\ref{tbl1}.

\begin{figure}
  \centering
  \includegraphics[width=0.85\textwidth]{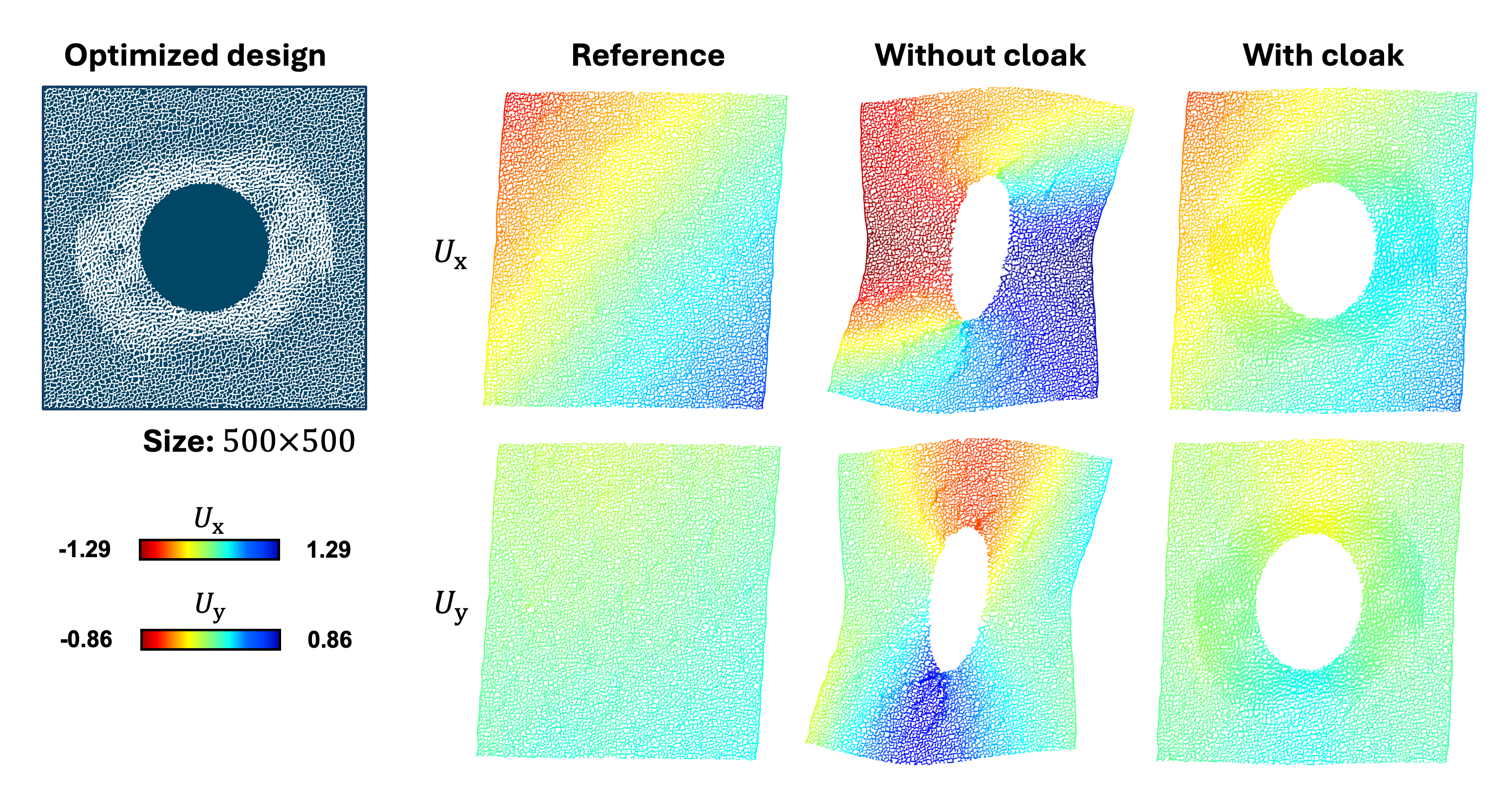}
  \caption{NCA-generated full cloaking structure at $500 \times 500$ resolution and its displacement field comparison. The left panel shows the generated heterogeneous microstructure in the given resolution. The right panels compare $U_x$ and $U_y$ across three configurations, including the reference structure, the uncloaked structure containing only the void, and the cloaked structure with the NCA-generated microstructure. The uncloaked structure produces severely distorted displacement fields, while the cloaked structure substantially recovers the smooth displacement distribution in the reference region, achieving a mean relative cloaking error of 26.12\% $\pm$ 1.71\% across 20 independently generated samples.}\label{fig_20}
\end{figure}

\begin{figure}
  \centering
  \includegraphics[width=0.85\textwidth]{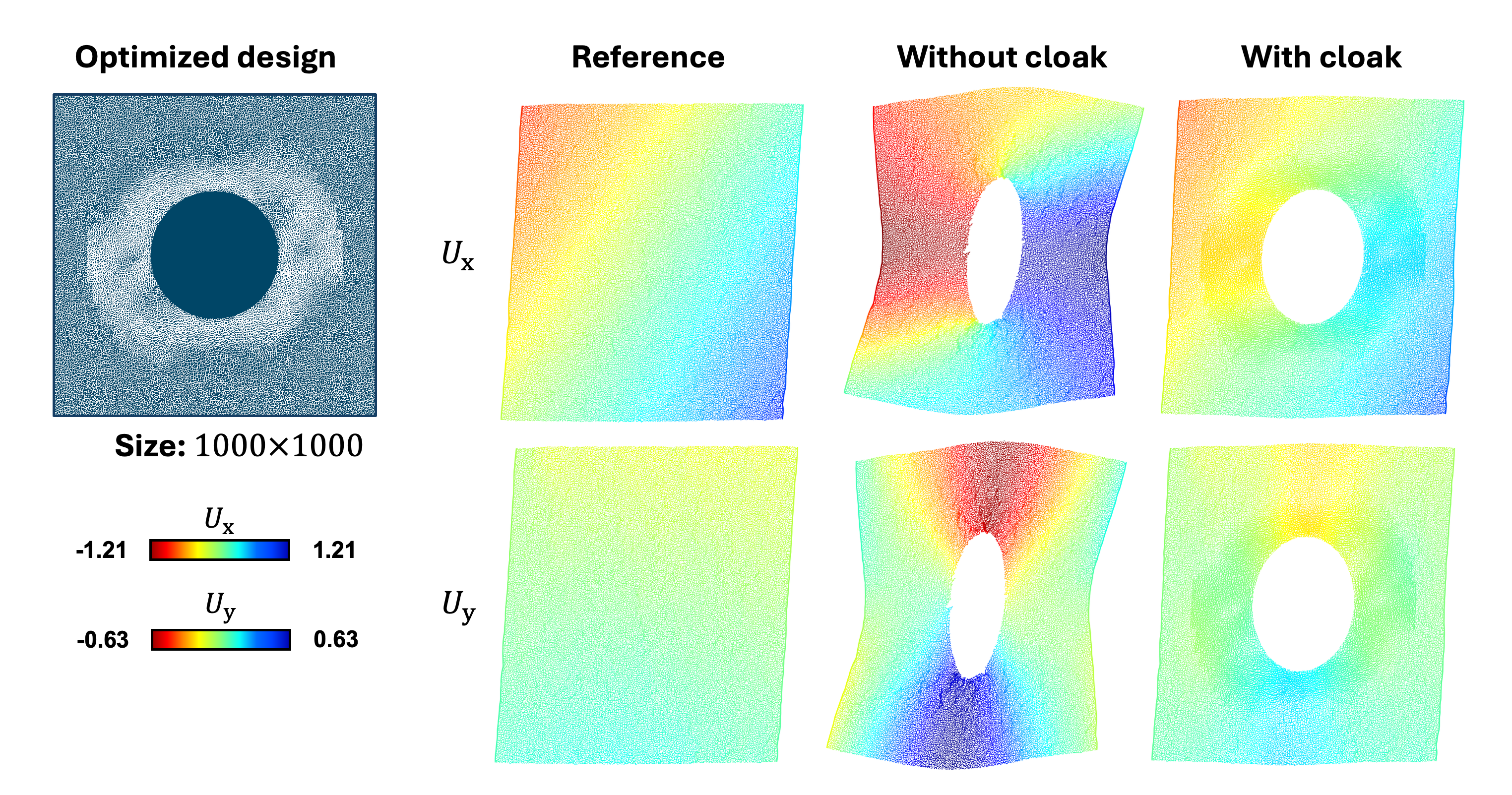}
  \caption{NCA-generated full cloaking structure at $1000 \times 1000$ resolution and its displacement field comparison, following the same layout as Fig.~\ref{fig_20}. The higher resolution produces smoother displacement fields and finer morphological detail in the generated microstructure, achieving a mean relative cloaking error of 16.86\% $\pm$ 0.63\% across 20 independently generated samples, demonstrating that the NCA framework achieves improved cloaking quality and consistency at higher resolutions with a single optimization.}\label{fig_21}
\end{figure}

Using these spatially varying control parameter fields, we then generate full-size cloaking structures with the trained NCA model. The control parameters act as growth cues that steer the self-organizing dynamics of the trained NCA, analogous to how local environmental cues guide natural growth. Because optimization is performed on properties rather than on any specific microstructure, the optimized design can be realized at any target resolution and on arbitrary discretizations, from regular grids to irregular meshes, without assembling discrete building blocks or retraining. As a demonstrative case, Figs.~\ref{fig_20} and~\ref{fig_21} show the generated structures and their displacement fields at $500 \times 500$ and $1000 \times 1000$ resolutions, respectively. In each case, we compare three configurations: the homogeneous reference structure, the uncloaked but voided structure, and the cloaked structure generated by the NCA. The void in the uncloaked structure strongly distorts both $U_x$ and $U_y$, leading to the largest relative error in Table~\ref{tbl1}. In contrast, the NCA-generated cloaked structure substantially restores the smooth displacement distribution of the homogeneous reference in the reference region. It can be noted that there are smooth morphological transitions, which are inherently guaranteed by the local self-organizing dynamics of the NCA, without requiring any post-processing steps. This provides a major advantage over existing building block assembly methods~\cite{WangGenerative2020}, which suffer from scalability and compatibility issues and cannot easily adapt to different discretizations or meshes.

The resolution study further demonstrates the flexibility of the generation framework. The $1000 \times 1000$ structure produces finer microstructural detail and smoother displacement fields than the $500 \times 500$ structure, with the mean relative displacement error over 20 independent samples decreasing from $26.12\% \pm 1.71\%$ to $16.86\% \pm 0.63\%$, indicating that higher resolution improves both the average cloaking performance and the consistency across independently generated samples. This improvement reflects the statistical nature of the disordered microstructures: the optimized effective properties represent the average response of a representative volume, which is perfectly recovered only when each design region contains enough features. Finer microstructures place more features within each region, reducing the property fluctuations and thereby improving both the accuracy and the consistency of the generated structure.
Even at $500 \times 500$ resolution, the NCA-generated structure still significantly reduces the displacement mismatch relative to the uncloaked case. These results confirm that optimized control parameter fields can be freely deployed across different scales, offering the flexibility to adapt the target resolution to practical manufacturing constraints.

\begin{table}
\caption{Cloaking performance with different configurations}\label{tbl1}
\begin{tabular*}{\tblwidth}{@{}LLLL@{}}
\toprule
\textbf{Configuration} & \textbf{Mean Rel. Error (\%)} & \textbf{Std (\%)} & \textbf{Samples} \\
\midrule
Uncloaked reference                  & 129.41  & ---    & ---  \\
Optimized result               & 8.10    & ---    & ---  \\
\midrule
NCA generated ($500 \times 500$)     & 26.12   & 1.71   & 20   \\
NCA generated ($1000 \times 1000$)   & 16.86   & 0.63   & 20   \\
\bottomrule
\end{tabular*}
\end{table}

\section{Conclusion}\label{sec:conclusion}

In this work, we developed an NCA-based generative framework for disordered microstructures. Instead of directly learning the geometry of microstructures, the framework learns the underlying local rules that grow them. This makes it highly data-efficient, requiring only a single template to enable microstructure growth across different domains and scales. The locality of the NCA update rule and its self-organizing nature ensure good connectivity and smooth transitions in spatially varying microstructures by design. This capability mirrors natural growth processes, where spatially varying cues enable heterogeneous growth while maintaining seamless integration across regions. A byproduct of this capability is that the NCA framework enables controlled asynchronous growth, allowing flexible control over where and when microstructures evolve. Similar to biological tissues, different regions can now grow asynchronously while maintaining seamless connectivity. This capability facilitates efficient parallel generation of microstructures over large and irregular domains and opens opportunities for adaptive and self-repairing material systems. These features distinguish the framework from existing building-block assembly methods, which require interface compatibility treatments and often suffer from scalability limitations.

An important insight from this study is that the learned NCA can be viewed as a generalized PDE system, allowing mature computational tools and analytical principles from the PDE domain to be leveraged for microstructure generation. Beyond simply replicating the template morphology, we enable continuous and interpretable control over the orientation, anisotropy, scale, and directional thickness of the generated microstructure by modifying the local perception operator, without additional training. This allows a single trained model to cover a broad morphological and mechanical property space that extends well beyond the training template, a capability that conventional deep generative models cannot easily achieve. Furthermore, by adapting only the local perception operator based on established principles from PDE discretization, the learned growth rule can be transferred across different domain geometries and representations, including planar and curved surfaces as well as regular grids and irregular meshes. This representation independence provides substantially greater flexibility than existing generative models and assembly-based approaches that are tied to predefined geometries and discretizations.

With these new capabilities enabled by the NCA, we extend beyond microstructure generation and integrate the trained NCA model into multiscale structural design. Demonstrated through a mechanical cloaking problem, this approach establishes an extensible pathway from a single microstructure template to spatially programmed heterogeneous structures, highlighting the potential of local self-organizing generative principles for next-generation material design.

The proposed framework lays the foundation for several promising future extensions.  First, while current controls steer the morphology and thereby influence the mechanical properties, conditioning target mechanical properties directly into the NCA model would extend the current framework to on-demand inverse generation of microstructures with target properties. Second, the framework can be extended beyond linear elastic properties to consider nonlinear mechanical responses, which will broaden the applicability of the framework to a wide range of mechanical behaviors that motivate disordered microstructure design. Third, the current 2D NCA can be extended to generate 3D microstructures by integrating 3D perception operators, such as 3D convolution kernels on voxel grids or point clouds.

We envision that the proposed framework will serve as a general generative engine for mechanical metamaterial design. Its one-shot efficiency, controlling flexibility and geometry adaptivity can help design in engineering domains that demand disordered microstructures with  prescribed mechanical properties, including biomedical implant design~\cite{Zhang2017, Cheikho2023, BabazadehNaseri2023, Poh2019} and soft robotics~\cite{Schaffner2018, Connolly2017, Ni2022}.








\appendix
\section{Model setups}
Table.~\ref{tbl_a1} provides the definition of hyperparameters in detail and the choice to train the NCA model. Table.~\ref{tbl_a2} provides the detailed setup for the optimization process.

\setcounter{table}{0}
\renewcommand{\thetable}{A\arabic{table}}

\begin{table}[pos=h]
\caption{Hyperparameters in the NCA training process}\label{tbl_a1}
\begin{tabular*}{\tblwidth}{@{}LLL@{}}
\toprule
\textbf{Hyperparameter} & \textbf{Value} & \textbf{Description} \\
\midrule
Hidden width \(D\)                     & 96                        & Width of the MLP hidden layer \\
Trainable parameters                   & 2{,}016                   & Parameters of \(\mathbf{W}_0\) and \(\mathbf{W}_1\) \\
Pool size \(N_{\mathrm{pool}}\)        & 256                       & Number of states in the sample pool \\
Batch size \(B\)                       & 4                         & States sampled per iteration \\
VGG feature layers                     & \(\{1, 6, 11, 18, 25\}\)  & Indices of extracted VGG-16 layers \\
Projection count \(P\)                 & 32                        & Random projections per sliced OT loss \\
Initial learning rate \(\eta_0\)       & \(10^{-3}\)               & Adam optimizer initial step size \\
Training epoch                        & 2{,}000                       & Number of total training epochs \\
LR decay factor                        & 0.3                       & Multiplicative decay at each milestone \\
LR decay milestones                    & 1{,}000                      & Iterations at which LR is decayed \\
\bottomrule
\end{tabular*}
\end{table}

\begin{table}[pos=h]
\caption{Detail setups in the multiscale optimization process}\label{tbl_a2}
\begin{tabular*}{\tblwidth}{@{}LLL@{}}
\toprule
\textbf{Parameter} & \textbf{Value} & \textbf{Description} \\
\midrule
Latent dimension $d_z$                  & 4                       & Dimension of VAE latent space per design element \\
Tube fraction $r_{\mathrm{frac}}$       & 0.05                    & Scaling factor for latent tube radius\\
Constraint limit $\gamma$               & 0.2                     & Allowable mean fraction of elements outside the tube \\
Spatial filter radius $r_{\min}$        & 5                       & Sensitivity filter radius over design elements \\
MMA damping                             & 0.01                    & Update weight for latent variables \\
\bottomrule
\end{tabular*}
\end{table}

\section{Supplementary videos}

\noindent\videolabel{vid:nca growth}: The growth process of leaf vein microstructure in Fig.~\ref{fig_2}.

\noindent\videolabel{vid:individual control}: Growth process under individual controls in Fig.~\ref{fig_6}, \ref{fig_8}, \ref{fig_9}, \ref{fig_11}.

\noindent\videolabel{vid:type transition}: Growth transitioning across multiple types of microstructure with irregular boundaries in Fig.~\ref{fig_14}(c).

\noindent\videolabel{vid:asynchronous growth}: Asynchronous growth in an irregular domain with customized growth sources and sequence in Fig.~\ref{fig_16}.

\noindent\videolabel{vid:bone repair}: In-situ bone repair process in Fig.~\ref{fig_17}.

\printcredits

\section*{Acknowledgments}
The authors acknowledge the support from the Department of Mechanical Engineering at Carnegie Mellon University.
\bibliographystyle{model1-num-names}
\newpage
\bibliography{cas-refs}



\end{document}